\documentclass[aps,pre,twocolumn,showpacs,amsmath,amssymb,
superscriptaddress]{revtex4}

\usepackage{graphicx}% Include figure files
\usepackage{bm}% bold math

\begin{document}

\newcommand{\be}{\begin{eqnarray}}
\newcommand{\ee}{\end{eqnarray}}

%\draft

\title{Spontaneous pinch-off in rotating Hele-Shaw flows}

\author{R. Folch}
\affiliation{Universiteit Leiden, Postbus 9506, 2300 RA Leiden, The
Netherlands}
\author{E. Alvarez-Lacalle}
\author{J. Ort\'\i n}
\author{J. Casademunt}
\affiliation{Departament ECM, Universitat de Barcelona, Av. Diagonal 647,
E-08028, Barcelona, Spain}

\begin{abstract}
The dynamics of the interface between
two immiscible fluids in a rotating Hele-Shaw
cell are studied experimentally, theoretically and by phase-field
simulations of the H-S equations using standard boundary
conditions.
As the central, denser fluid is
centrifuged, it forms fingering patterns with long, thin radial
filaments ended by a droplet, alternating with incoming
fingers of the outer, less dense fluid. Simulations show the
length (width) of the filaments to grow (decay) roughly
exponentially, and
the incoming finger tips to asymptotically approach a finite radius
for $n$-fold symmetric initial conditions; these thus tend to a
stationary-shape ``star-fish'', whose form is calculated.
The filament width decays with a time constant which
depends only on the viscosity contrast, whereas its
length exhibits
a completely universal growth rate; the latter
is related to the run away of an isolated droplet,
for which we give an exact solution.
The exponential behavior is clear for high viscosity contrasts $A$,
but important deviations are found for low $A$.
Both experiments and simulations show systematic pinch-off of the
droplets at the tips of the filaments for low and not for high $A$.
A lubrication approximation is derived and successfully accounts for the
filament thinning and the differences with $A$; in particular, it explains why
pure exponential thinning is not observed for low $A$, and it could clarify the
presence or absence of finite-time pinch-off itself, since the agreement of
experiments and simulations suggests that this phenomenon is contained in the
Hele-Shaw equations.
This agreement includes both high- and low-$A$ morphologies, and
the growth rate of the filament length for high $A$; for low $A$,
the experimental time constant appears to be different from that
predicted by standard Hele-Shaw boundary conditions and observed
in simulations. An effective slip condition for the Poiseuille
flow of inner liquid across the cell gap in the case of two
liquids gives a possible explanation of this discrepancy.

\end{abstract}


\maketitle

\section{Introduction}
\label{intro}

Topological singularities such as interface pinch-off in fluid
flows have been the object of intense study in the last decades
\cite{scalefree,Eggers97,Science00,Eggers02,Oron97,Zhang99,Lee02,
Goldstein93,Duppont93,Constantin93,Goldstein95,Goldstein98,
Almgren95,Almgren96}. The pinch-off of a droplet from a volume of
fluid embedded in another one is easy to observe in everyday life,
and has fascinated both physicists and applied mathematicians.
Because the interface between the two fluids shrinks to
arbitrarily small scales at the point of detachment, a continuum,
coarse-grained description must break down at some stage. The
failure of the macroscopic description is reflected in the
spontaneous generation of singular behavior. In the neighborhood
of such singularities, the problem might become scale-free, giving
rise to self-similar scaling behavior, for which some degree of
universality is naturally expected \cite{scalefree}. Interestingly,
hydrodynamics alone have been found sufficient to account for the
occurrence of finite-time pinch-off in several cases. 
It is remarkable that the continuum, hydrodynamic description 
may correctly predict the occurrence of finite-time singularities 
out of a smooth initial condition, 
and properly describe them all the way to the
very microscopic scales where atomic-scale forces take over. These
forces then implement the interface breakup and reconnection.
This is the case, for instance, of
three-dimensional jets \cite{Eggers97}.
The pursuit of this idea in the specific context of rotating
Hele-Shaw flows is the basic motivation of the present study.

Recently, macroscopic hydrodynamic equations have also been used
as a basic ingredient in the dynamics of nanojets
\cite{Science00,Eggers02}, or in very thin membranes, where the
effects of Van der Waals forces must be taken into account
\cite{Oron97,Zhang99}. These are examples of nonlinear processes
where the macroscopic dynamics coupled with molecular forces or
thermal noise lead to pinch-off phenomena.

In the narrow gap between the two parallel glass plates of a
Hele-Shaw cell, the scale where the effective two-dimensional
``macroscopic'' description fails is basically the thickness of the
cell gap \cite{Lee02}. Although this cutoff is much
larger (typically, of the order of the millimeter) 
than the microscopic cutoff of
the hydrodynamic description of 3d jets, it still makes sense to
investigate to what extent the 2d effective
dynamics in a Hele-Shaw cell (playing the role of the ``macroscopic''
dynamics) lead to the spontaneous occurrence of finite-time
singularities. The interface recombination will
certainly introduce new physics (not contained in the ordinary
Hele-Shaw equations) which could easily depend on details that
may not be universal, such as wetting conditions, 3d structure of
the meniscus, contact line motion, etc.; the precise time of the
pinch-off may also depend on these details. However, the underlying idea
is that the effective cutoff and the details of the
actual 3d pinch-off can in principle be arbitrarily reduced, for
instance changing the cell gap. So, ultimately, the relevant
question is what the prediction of the effective 2d
(Hele-Shaw) dynamics is: If the 2d dynamics lead themselves to
finite-time pinch-off, this would certainly assure the existence
of finite-time pinch-off in the real experiments, and would also
define an upper bound for the actual experimental pinch-off. If,
on the contrary, no finite-time pinch-off is obtained for the
Hele-Shaw dynamics, it will be relevant to study the evolution of
a thin neck as predicted by the effective 2d dynamics, in order to
elucidate, in each particular case, whether actual pinch-off would
be expected invoking the additional 3d effects. In this context,
the simulations performed within a phase-field scheme are
particularly appropriate, because they incorporate a natural cutoff
(the interface thickness) which controls the actual pinch-off and
which, like the cell gap in Hele-Shaw flows, can be modified
at will to postpone the pinch-off of a narrow fluid filament.

 It is well known that the effective two-dimensional Hele-Shaw
dynamics can lead {\it per se} to pinch-off. In fact, it was for
this type of dynamics that finite-time pinch-off singularities in
a hydrodynamic description were first found \cite{Goldstein93,
Duppont93}. As opposed to the 3d case of a cylindrical interface, 
a 2d fluid filament (two parallel interfaces) 
is stable to the Rayleigh criterion:
The total interface length (at constant area) is larger for any 
perturbation around straight interfaces. 
However, surface tension alone has been shown to
drive a configuration of two droplets of fluid
connected by a neck to finite-time pinch-off in two-dimensional
simulations, for certain initial conditions \cite{Almgren95,
Almgren96}. A straight filament of fluid can also be made to pinch
at infinite time (and sometimes also at finite time), with very
specific boundary conditions
\cite{Goldstein93,Duppont93,Constantin93,Goldstein95,
Goldstein98}. However, two-dimensional pinch-off in unprepared
situations emerging spontaneously in externally driven dynamics,
such as in traditional viscous fingering experiments, has not been
addressed. One of our main goals is to account for the {\em
spontaneous} approach to pinch-off singularities often observed
while a highly nonlinear pattern develops following a
morphological instability, in particular for rotating Hele-Shaw
flows. Pattern formation and pinch-off singularities can indeed be
intimately related, as it will become apparent in the case studied
here.

Another issue that it is worth addressing is the role of the viscosity
contrast or Atwood ratio $A\equiv (\mu_{\rm in} - \mu_{\rm out})/
(\mu_{\rm in} + \mu_{\rm out})$  (where $\mu_{\rm in}$ and
$\mu_{\rm out}$ are the viscosities of the inner and outer fluids
respectively) in the pinch-off process. It is known that this
parameter has a strong influence on the interface shape near
pinching in three dimensions \cite{Zhang99}, but only recently has
it been remarked that the most studied limit $A\to 1$ might be a
very special case also in three dimensions \cite{Sierou03}. In
effectively two dimensions, experiments performed in this limit,
with air displacing a liquid in a channel geometry, show that the
fingers formed due to the morphological instability compete until
a single finger is left \cite{st}; the neck of the transient and
final finger(s) does not even pinch. In contrast, when a denser
liquid is put on top of a less dense one of similar viscosity
(typically $A \sim$0--0.5) in the same but tilted channel, 
fingers do not compete, but elongate to
form thin filaments with a droplet at their tip \cite{maher},
which can indeed pinch off \cite{footnote0}.
One possible scenario to explain this different tendency to pinch-off
might consist in relating the absence of competition for low $A$ with the
formation of long filaments, which could then pinch by mechanisms
not qualitatively different from those of previous studies
\cite{Goldstein93,
Duppont93,Almgren95, Almgren96, Constantin93,Goldstein95, Goldstein98}.

We have performed experiments in a rotating
Hele-Shaw cell, where a drop of the more dense fluid is placed at
the center and the instability is driven by the centrifugal force.
The idea is to favor the formation of long (radial) filaments by
the use of a driving force which increases as the fingers become
longer. The natural question to pose is then whether these long
filaments will also pinch for high viscosity contrasts.

Long fingers are indeed formed and they stretch and narrow for any
viscosity contrast $A$. However, for high values of $A$ we only
observe pinch-off sporadically, whereas we find it systematically
for low ones. This rules out the above scenario. It
rather suggests that the precise nonlinear dynamics close to pinch-off, 
and not only the overall morphology, is also dependent on the
viscosity contrast $A$. This dependence could either be contained
in the (2d) Hele-Shaw dynamics or, instead, enter through
different wetting properties for air-oil (high $A$) and oil-oil
(low $A$) interfaces when the width of the filaments becomes
comparable to the cell gap in our experiments.

We have run phase-field simulations of the two-dimensional dynamics
which lead to pinch-off singularities for low, but not high
viscosity contrast. Although we can only track the first stages of
the approach, the major dependence of the pinching dynamics on the
viscosity contrast thus seems to be contained in the Hele-Shaw
equations. In particular, our simulations indicate
that finite-time pinch-off could be contained (at least) in
low-$A$ Hele-Shaw dynamics for more general settings than those
originally studied in Refs.
\cite{Goldstein93,Duppont93,Almgren95, Almgren96,Constantin93,Goldstein95,
Goldstein98}. Therefore, it is plausible that the effect of the
cutoff (here, the cell gap), 
although entering the problem earlier than in usual
three-dimensional cases, can still be regarded as an
implementation of pinch-off and reconnection, not necessarily
affecting significantly the approach to pinch-off, which is
governed by the Hele-Shaw dynamics.

The rest of the paper is organized as follows: In Sec. \ref{exp}
we present the experimental results, with an indication of scaling
behavior. This is then explained in the following section by
simple theoretical arguments valid asymptotically. In Sec.
\ref{numerics} we test this basic picture by numerical simulations
of the Hele-Shaw equations. These confirm theory and experiments
to agree for high viscosity contrasts, whereas puzzles remain
for low ones. One of these can be solved by rethinking the
experimental time scale as done in Sec. \ref{sectimescale}, 
while another requires a refinement of the
theory which is performed in Sec. \ref{lubr}.
The results are summarized and further
discussed in Sec. \ref{discussion}.

\section{Experimental results}
\label{exp}

We have performed two sets of experiments in a rotating Hele-Shaw
cell: one for high and one for low viscosity contrast. These two
types of experiments were already performed for other purposes in
\cite{Alvalow03} and \cite{Alvahigh03}; we repeat them here to
address the scaling behaviors and the presence of pinch-off as
functions of viscosity contrast. Our cell consists of two
horizontal glass plates, $7$ or $10$ mm thick and $390$ mm in
diameter, separated by different spacers of height $b=0.5, 0.7,
0.8, 1.0, 1.4$ mm. The cell is rotated around a vertical axis
intersecting its center, with a controlled frequency $\Omega$, and
the interface shape is recorded with a CCD camera.

In a first set of experiments, a silicone oil of viscosity
$\mu_{\rm in}=50$cp and density $\rho_{\rm in}=975\pm10$kg/m$^3$
at $20^\circ$C and air are used as inner an outer fluids,
respectively, in a prewetted cell \cite{prewet}. This yields $A=
1$ and $\Delta\rho=\rho_{\rm in}-\rho_{\rm out}\simeq
0.98$g/cm$^3$, and an interfacial tension $\sigma=20.7$ mN/m. In a
second set of experiments, we use another silicone oil ($\mu_{\rm
in}=530\pm50$cp, $\rho_{\rm in}=975\pm10$kg/m$^3$) as inner, and a
vaseline oil ($\mu_{\rm out}=190\pm50$cp, $\rho_{\rm
out}=875\pm10$kg/m$^3$) as outer fluid. In this case $A=0.45\pm
0.05$, $\Delta\rho\simeq 0.10$ g/cm$^3$, $\sigma=1.8\pm0.7$mN/m.
Mean values are given at $20^\circ$C, and the uncertainty in the
viscosities (and their contrast $A$) accounts for temperature
variations. 

\begin{figure}
\begin{center}
\includegraphics[width=8 cm]{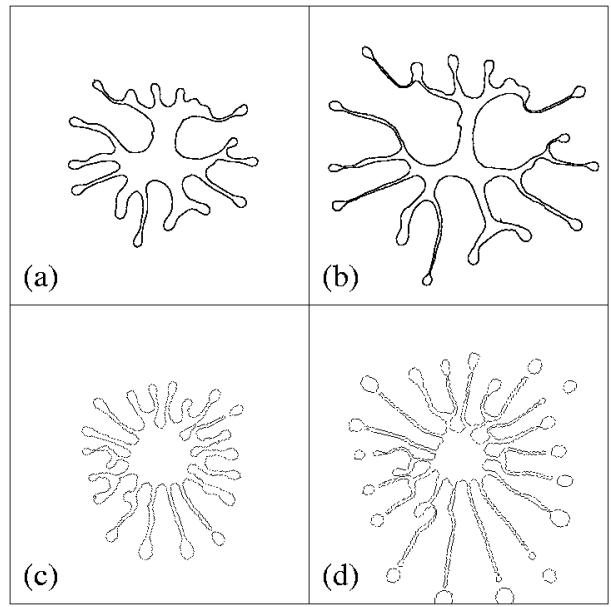}
\caption{Pattern evolution for $A= 1$, $b=0.5$mm,
$\Omega=120$rev/min and $R_0=50$ mm (a,b), and
$A\simeq 0.45$, $\Omega=180$rev/min and $R_0=38$ mm (c,d).
Snapshots ($3R_0\times 3R_0$) 16.5s (a) and 22.5s (b),
and 122s (c) and 158s (d) after beginning rotation.}\label{morfexp}
\end{center}
\end{figure}

Figures \ref{morfexp}(a,b) and \ref{morfexp}(c,d) show the typical
patterns formed at two different stages, for high ($A= 1$) and low
($A\simeq$0.45) viscosity contrast, respectively. Although the
nonlinear dynamics depend significantly on the viscosity contrast,
as we can see, let us focus for the moment on those features which
are generic to any value of $A$: After some latency time, the
linear instability leads to small undulations on the initial
circle of radius $R_0$. These undulations grow into radial fingers
of silicone oil. As they are further centrifuged, the fingers
develop overhangs; their tips evolve into differentiated droplets,
whereas the overhang regions themselves narrow and stretch to
become thin filaments, which continue to stretch and narrow. In
parallel with this, incoming fingers of less dense fluid advance
towards the cell center. However, they do not develop thin
filaments nor droplets, and they slow down as they approach the
cell center.

\begin{figure}
\begin{center}
\includegraphics[width=7 cm]{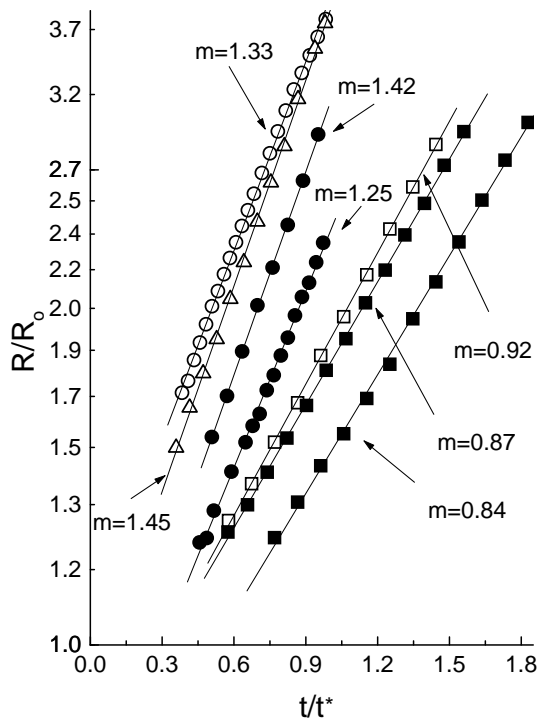}
\caption{Droplet kinetics: Evolution (as a function of scaled
time) of the radial coordinate of (i) the center of mass of the
drop at the tip of three different filaments for $A\simeq 0.45$,
before and after pinch-off (circles); (ii) the center of mass of an
isolated drop for $A\simeq 0.45$ (triangles); (iii) the tip of
three different filaments for $A=1$ (squares). Circles and
triangle have been translated to earlier times for comparison.
Open symbols correspond to the most excentric drops in Fig.
\ref{morfexp}; $m$ are the slopes of linear
fits.}\label{stretchingexp}
\end{center}
\end{figure}

The filament thinning and the dynamics of the incoming fingers are
rather difficult to measure accurately and not very sharply
defined experimentally. In contrast, the stretching of an
individual filament, or, more precisely, the radial coordinate of
the droplet at its tip, $R$, is more readily accessible. The
squares ($A= 1$) and the circles ($A\sim 0.45$) in Fig.
\ref{stretchingexp} indicate this droplet position (even after
pinch-off) in units of the radius $R_0$ of the initial circle and
in log scale, as a function of time $t$ in units of the time scale
\be \label{timescale} t^*=\frac{12(\mu_{\rm in}+\mu_{\rm out})}
{b^2\Omega^2\Delta\rho}. \ee
The open squares and circles indicate
the evolution of the furthest droplet in Figs. \ref{morfexp}(a,b)
and \ref{morfexp}(c,d) respectively.

As we can see, the growth tends to be roughly exponential, at
least for $R/R_0>2$. (Note that $R/R_0<2$ corresponds rather to
relatively small fingers, rather than droplets at the tip of thin
filaments). It seems reasonable to measure a growth rate
---the slope of linear fits (solid lines) in Fig. \ref{stretchingexp};
this yields $m=0.8$--1.0 with good linearity beyond
$R=2R_0$ for $A=1$, and $m=1.25-1.5$ for $A\simeq 0.45$.
$A\simeq 0.45$ curves tend to bend a bit more, but this
eventual bending is uncorrelated with the droplet pinch-off: To
check this, we have performed an experiment with an isolated,
off-center circular droplet; it roughly keeps its circular shape and runs
away at a very similar rate ($m=1.45$, triangles).

The most striking difference between high ($A=1$) and low
($A\simeq 0.45$) viscosity contrast is the systematic droplet
pinch-off observed for low, but not high $A$. For the latter,
filaments keep on growing and stretching, while reaching a width
comparable to the cell gap, that is close to the natural cutoff.
While we have not spanned the whole range of $A$, all previous
evidence shows that the so-called low contrast behavior is
qualitatively unchanged for most of the range of $A$, and only
very close to $A=1$ significant differences are found. This has
been recently discussed in Refs. \cite{Paune-thesis,Alvalow03}.

Another difference seems to be the growth rate of droplets (connected
or not to filaments). The uncertainties in the viscosities are not sufficient
to account for it: droplets seem to be centrifuged faster in
dimensionless time for low than for high viscosity contrasts.
Our discussion on this is postponed to Sec. \ref{sectimescale}.

\section{Theoretical scenario}
\label{basics}

We try to link here by continuity arguments the filament thinning and
stretching and the evolution of incoming fingers, and calculate the
various growth and decay rates in a certain approximation. Because the
dynamics of droplets at the tip of filaments should affect their
stretching,  we also construct an exact time-dependent solution for an
off-center isolated droplet. Finally, we explore the existence of
stationary-shape solutions as possible asymptotic attractors suggested
by the slowing down of incoming fingers.

\subsection{Scalings}
\label{scalings}

Fluid incompressibility imposes fluid area conservation, which
intuitively forces filaments to stretch and/or feed the droplet at
their tip as they narrow, and fingers of less dense fluid to grow
inwards as mass is transported outwards through the filaments.
Area conservation hence seems an important feature of the dynamics. In
particular, ``star-fish'' patterns have been observed in an {\it ad
hoc} geometrical model for the normal velocity of a front with
{\em local} mass conservation on the interface (+ a constant flux)
\cite{geometrical}. However, incompressibility also generally
leads to {\em non-locality}, which will turn out to be crucial for
low viscosity contrasts and for the presence of pinch-off
singularities.

\begin{figure}
\begin{center}
\includegraphics[width=0.45\textwidth]{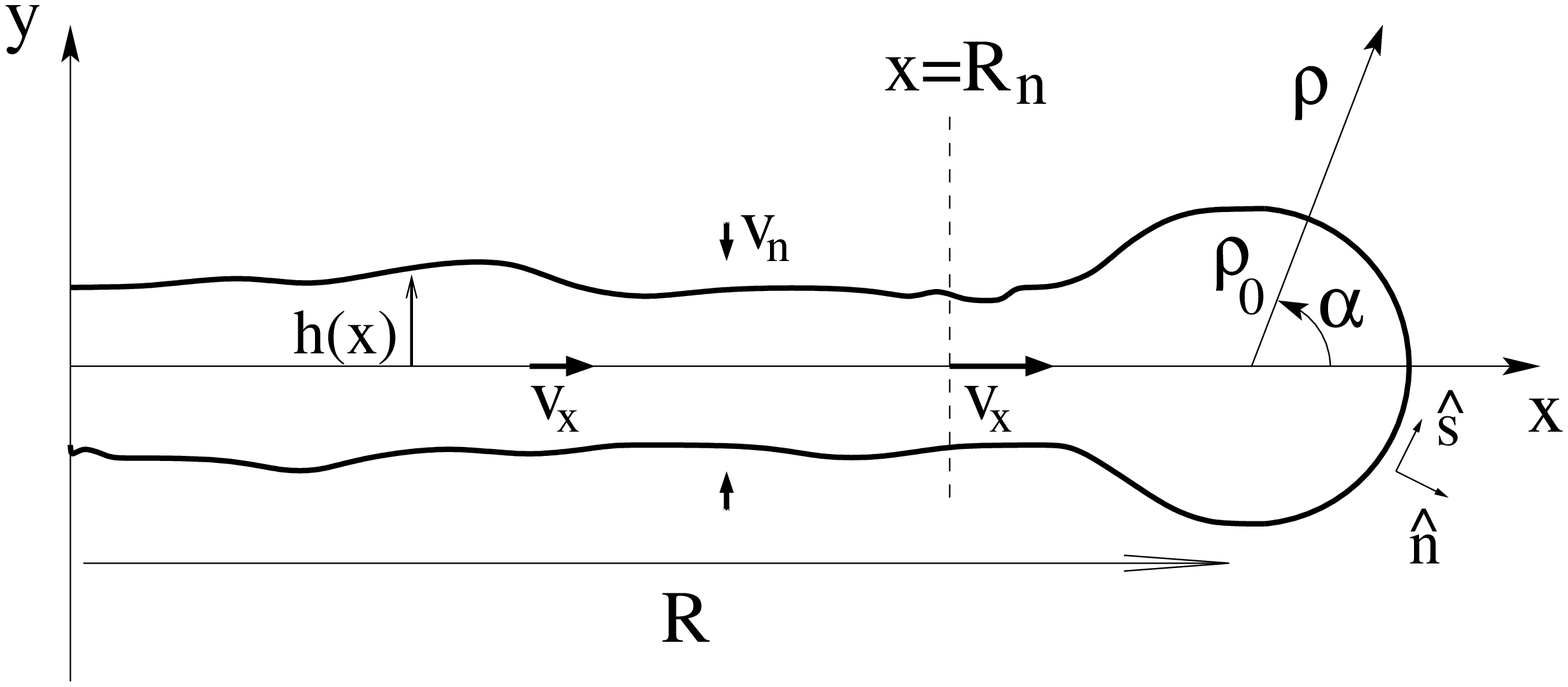}
\caption{Sketch of a filament of fluid with a droplet at its tip.
}\label{sketch}
\end{center}
\end{figure}

Let us first address the filament dynamics by means of
a continuity equation in the overall
direction of the filament $x$, which reflects the
incompressibility of the inner fluid:
\be
\label{hcontinuity}
\partial_t h+\partial_x (hv_{\hat{x}})=0,
\ee where $h(x,t)$ is the local y coordinate of either the upper
or the lower interfaces, and $v_{\hat{x}}$ is the $x$ component of
the fluid velocity inside the filament (see Fig. \ref{sketch}).

Motivated by the experimental observations, we look for a scaling
solution of the form
\be \label{scalansatz}
h(x,t)=L(t)h_0[xL(t)], \ee
where $L(t)$ is a dimensionless scaling factor. Substituting this
ansatz into Eq. (\ref{hcontinuity}), we find
\be
\label{varsep} -\frac{\partial_t L}{L} = \frac{h_0 \partial_x
v_{\hat{x}} + v_{\hat{x}}\partial_x h_0} {h_0+x \partial_x h_0} =
K, \ee where $K$ is independent of $x$.

If $v_{\hat{x}}$ happens to be either (i) a constant along the
filament or (ii) proportional to $x$, $v_{\hat{x}}=Kx$, a
perfectly straight filament ($h_0$ independent of x) satisfies Eq.
(\ref{varsep}) and hence stays straight. Moreover, if and only if
$v_{\hat{x}}=Kx$, {\em any} function $h_0(x)$ satisfies Eq.
(\ref{varsep}) and preserves its form. Interestingly, we will see
that, for a straight filament, (i) corresponds to the standard
channel geometry, and that our setup mostly falls in case (ii). In
contrast, the radial case with injection, for example, cannot
exhibit this type of self-similar solutions.

We find that, whenever the above scaling ansatz is valid, and
further assuming $K$ to be constant in time, then
%
%\be
%\label{exponential}
$L=e^{-Kt}$.
%\ee
%
The initial condition $h_0(x)$ thus narrows exponentially
\begin{equation}
\label{scalingh}
h=h_0e^{-K t}
\end{equation} and stretches
\begin{equation}
\label{scalingr}
R_{\rm n} = R_{\rm n}(t=0)e^{K t}
\end{equation}
(where $R_{\rm n}=x\;|\;\partial_t h_0(x,t)=0$)
with a time constant set by the inner fluid velocity. Figure
\ref{sketch} schematically shows the in- and outgoing flow in
this case.
Eq. (\ref{scalingr}) is a statement that a reference point on the
interface is advected with the filament as this is stretched.
In particular, if applied to the neck of the filament just before the
droplet as represented in Fig. \ref{sketch}, it would predict the total
filament length to grow with the same time constant $K$ with which
its width decays. However, note
that the outgoing flux of inner fluid at the neck can go into the
droplet or contribute to the filament stretching. If some fraction
of it actually goes into the droplet (as observed in the experiments),
the scaling of the radial coordinate of the neck will be delayed
with respect to the thinning rate $K$.

Finally, in order to relate this filament scaling with the
incoming fingers, consider a reference circle of radius $R_{\rm
ref}$ intersecting all filaments near their root, where they begin
to look straight. The outwards flux of denser fluid through the
circle and inside each filament is $2h(t)v_{\hat{x}}(x=R_{\rm
ref})$, and assuming again that $v_{\hat{x}}(R_{\rm ref})$ does
not depend on time, we find it to scale as $e^{-K t}$. In general,
the area of denser fluid enclosed by this reference circle $a$
should hence asymptotically decay with the same rate,
\begin{equation}
a=a(t\rightarrow \infty) +\left [a(t=0)-a(t\rightarrow
\infty)\right ]e^{-K t}. \label{scalingarea}
\end{equation}
A similar slowing down can then be expected for the radial
position of the tips of identical incoming fingers.

\subsection{Exponential behavior and time constant}
\label{exponential}

We now show that, {\em for a straight filament}, $v_{\hat{x}}$ is
linear in $x$ as we anticipated, and we find the time constant
$K$, which we have seen to set the time scale of the various
exponential behaviors discussed so far. For that purpose, let us
compute the tangential velocity jump across the interface.

The 2d velocity field $u_i$ is proportional to the gradients of the
pressure field $p_i$ (Darcy's law), where the subscript $i$ labels
the inner ($i=$in) and outer ($i=$out) fluids. For a rotating cell
\cite{Alvalow03}
\begin{equation}
\label{darcy}
\vec u_i=-\frac{b^2}{12\mu_i}
\left ( \vec\nabla p_i - \rho_i \Omega^2 r \hat r \right ),
\end{equation}
where $r$ is distance to the rotation axis and $\hat r$ is a unit
radial vector pointing outwards. Using the standard boundary
condition of  Laplace law for the capillary pressure jump across
the interface, $p|_{\rm in}-p|_{\rm out} = \sigma \kappa$, where
the bar $|$ stands for the limit value when approaching the
interface from either side and $\kappa$ is the interface
curvature in the cell plane, we obtain
\begin{eqnarray}
\nonumber
\hat s \cdot (\vec u|_{\rm out}-\vec u|_{\rm in}) &=&
\frac{b^2}{6(\mu_{\rm in}+\mu_{\rm out})}
\left ( \sigma\partial_s\kappa
-\Delta\rho\Omega^2 r \hat r\cdot\hat s \right )\\
\label{veljump}
&&+A \hat s \cdot \left (\vec u|_{\rm in}+\vec u|_{\rm out} \right ),
\end{eqnarray}
where $s$ is a coordinate tangential to the interface and $\hat s$
the unit vector in that direction.

In our sketch (Fig. \ref{sketch}), we identify $v_{\hat{x}}\simeq
\vec u|_{\rm in}\cdot\hat s$, which is a function of the local
interface geometry and $(1-A)\vec u|_{\rm out}\cdot\hat s$, the
solution of a non-local problem. To keep the problem local, we now
assume that the outer fluid approaches the filament normal to it,
$\vec u|_{\rm out}\perp\hat s$ (Note that this assumption is not
necessary for $A=1$).

For a straight ($\kappa=0$) radial ($\hat x\parallel \hat r$)
filament, we obtain the anticipated $v_{\hat{x}}=Kx$ and the time
constant
\begin{equation}
\label{timeconstant}
K= \frac{b^2\Delta\rho \Omega^2}
{6(\mu_{\rm in}+\mu_{\rm out})(1+A)} =
\frac{2}{1+A}\frac{1}{t^*}=\frac{b^2\Delta\rho \Omega^2}
{12\mu_{\rm in}}
\end{equation}
[case (ii) above]. One might wonder how a scale-free solution is
possible without requiring the filament width to be much smaller
than any other length scale, as it is legitimate to do close
enough to pinch-off \cite{scalefree}. The answer is that the
assumption of flatness has removed the capillary length
(proportional to $\sigma$) from the {\em local} problem; it still
sets, together with the radius of the initial condition $R_0$, the
size of e.g. the droplet at the tip of the filaments, but no
external length scale can be felt when dropping the nonlocal term
$(1-A)\vec u|_{\rm out}\cdot\hat s$. This allows for scaling
solutions to exist well before pinch-off for $A\equiv 1$. It is
this approach to pinch-off, rather than the pinch-off phenomenon
itself, which will be further studied in this article.

For a tangential filament, $K=0$
and the filament is stationary [case (i) above],
which is consistent with our observations for the tangential segments
of some filaments. These segments are found where a tip split
(for $A=0$, see Fig. \ref{morfexp}(c,d))
or at the point where an
incoming finger was overcome and stopped by its neighbors
(for $A=1$, see Fig. \ref{morfexp}(a,b)).

For the case of viscous fingering in a channel, the linear
increase in the centrifugal force is replaced by a constant
gravity (or equivalent injection). We can easily track this
difference to yield a fluid velocity inside a filament parallel to
the gravity or injection direction $v_{\hat{x}}$ independent of
$x$. The filament width will again be stationary [case (i) above]. We
see by comparison that rotation not only ensures the formation of
long filaments as we expected, but also brings the two interfaces
close together.

For not completely straight filaments $h(x)$, $v_x\neq Kx$.
Therefore, strictly speaking, the scaling solution Eq.
(\ref{scalansatz}) does not hold for any $h_0(x)$, but only for a
straight filament. It remains an approximate solution for slightly
curved filaments (see Sec. \ref{lubr}), but it makes little sense
for the droplets at the tip of the filaments. This raises the
issue of dynamics of the droplet, which could potentially affect
the filament stretching.

\subsection{Isolated droplet}
\label{droplet}

The velocity found for the filament stretching has an elementary
interpretation: In the overdamped limit associated to Darcy's law,
all forces balance. The inner fluid velocity $v_x$ will hence be
that which makes friction balance the other external forces,
namely the centrifugal force and the force exerted by the outer
fluid through its pressure gradients. The latter can be derived
from another force balance, now in the outer fluid. There, the
viscous forces vanish in the $x$ direction with our assumption
$\vec u_{\rm out}\perp\hat s\simeq\hat x$, or even without this
assumption for $A=1$, since it corresponds to $\mu_{\rm out}=0$.
Thanks to this, the other forces in the outer fluid (centrifugal
and pressure gradients) balance in this direction. Thus, back
in the inner fluid, friction, $12\mu_{\rm in}v_x/b^2$, only needs to
compensate for centrifugal forces, $\Delta\rho \Omega^2 x$. This
gives rise to the velocity found, $v_x=Kx$ with $K$ given by Eq.
(\ref{timeconstant}).

For a droplet (a
closed interface), in contrast,
the outer fluid needs to be moved in the $x$
direction. The outer viscosity and velocity field enter the force
balance, and $v_{\hat{x}}\leq Kx$, where $v_{\hat{x}}$ is now the
velocity of the center of mass of the droplet, and the equality holds
only for $A=1$. Except for $A=1$, it is clear that the filament
stretching will indeed be affected by this lower droplet velocity.
Furthermore, $v_{\hat{x}}$
will depend on the outer flow field, and hence on the droplet shape.

For a slightly off-center circular droplet, $v_{\hat{x}}$ is
known: Such a droplet corresponds to the mode 1 in a linear
stability analysis of a circular, centered drop. Its growth rate
is $K'=1/t^* =b^2\Delta\rho \Omega^2/[12(\mu_{\rm in}+\mu_{\rm
out})]$ \cite{Alvalow03}, where we readily see that it is the sum
of viscosities, and not only the inner one, what counts.

We now compute the velocity for a circular droplet $\partial_s
\kappa=0$ rigorously. Preserving the circular shape requires that
$\vec u_{\rm in}=\dot{R} \; \hat x$ everywhere ($R$ is distance
from the rotation axis to the center of the droplet), so that the
circle be just translated as a whole. We express the unit vectors
tangential ($\hat s$) and normal ($\hat n$) to the interface and
the radial vector $\vec r$ connecting the rotation axis and a
point on the interface in polar coordinates ($\rho$, $\alpha$)
with respect to the center of the droplet (see Fig. \ref{sketch}),
and compute $\hat s \cdot \vec u|_{\rm in}=-\dot{R} \sin\alpha$
and $\vec r\cdot\hat s = -R \sin\alpha$. For $A=1$, Eq.
(\ref{veljump}) reduces to a simple geometric relation between the
two. Interestingly, like the filament scaling, this is verified
either for a channel under injection or gravity ($\dot{R}$ is then
a constant) or for a rotating cell as presented here,
$\dot{R}=K'R$, but not for a general driving force or geometry. We
hence find that a circular off-center droplet is a time-dependent
solution in a channel or rotating cell for $A=1$ as was previously
found using conformal mapping techniques in \cite{tesifrancesc}.

For $A<1$, the outer flow enters the problem. If a circular
droplet is to remain a solution, it has to match the linear
regime. At that stage, we will still have $\dot{R}=K'R$,
regardless of the viscosity contrast. Substituting this into Eq.
(\ref{veljump}), we find $\hat s \cdot \vec u|_{\rm out}= -\hat s
\cdot \vec u_{\rm in}=\dot{R} \sin\alpha$. Continuity of normal
velocities across the interface gives us $\hat n \cdot \vec
u|_{\rm out}=\dot{R} \cos\alpha$, and we thus find the outer
velocity on the interface to be $\vec u|_{\rm out}=\dot{R}(\hat x
\cos 2\alpha + \hat y \sin 2\alpha)$. We then propose the outer
velocity field to be the product of its value on the interface and
a function of $\rho$ only. We find that this can indeed fulfill
incompressibility if the droplet is isolated, i.e., if there are
no other boundary conditions to be satisfied elsewhere. For the
channel geometry, this corresponds to the limit in which the
droplet is much smaller than the distance between the walls. The
complete solution reads
\begin{subequations}
\label{dropletflow}
\begin{eqnarray}
%\nonumber
\vec u_{\rm in} &=& \dot{R} \hat x, \\
\label{dropletfield}
\vec u_{\rm out} &=& \dot{R} (\rho_0/\rho)^2
(\hat x \cos 2\alpha + \hat y \sin 2\alpha),\\
\label{runaway}
R &=& R(t=0)e^{K't}.
\end{eqnarray}
\end{subequations}
The corresponding streamlines in the outer fluid $\rho/\rho_0
\propto \sin(\alpha)$, are circles of different radii tangential
to $y=0$ and passing through the center of the droplet, although
they change into horizontal lines inside it (see Fig. \ref{stream}
bottom right).

Hence, an off-center circle turns out to be an exact
time-dependent solution for any viscosity contrast and independent
of it, even beyond the linear regime. For $A=1$, $K'=K$ and the
filament stretching and the droplet runaway coincide regardless of
the droplet shape and whether it is near a filament or not. This
single time constant should hence be robust, and indeed an only
slightly lower value [($0.8$--1.0)$/t^*$] is systematically
observed in our experiments (squares in Fig. \ref{stretchingexp}).
For $A<1$, in contrast, $K'<K$, and an isolated, circular droplet
has a time constant down to twice smaller (for $A=0$) than that of
the filament stretching. For a droplet connected to the tip of a
filament (neither isolated nor circular), we do not know its
dynamics rigorously.
Our experiments are puzzling in this respect:
Isolated droplets (triangles in Fig. \ref{stretchingexp})
and those connected to a filament (circles)
display the same growth rate;
consistency with the theory then requires it to be $K'$ (the same
than for $A=1$), since the isolated droplet solution is exact.
However, we seem to find a growth
rate of $K$ [($1.25$--1.5)$/t^*$],
and, in any case, higher than the measured
value for $A=1$.
Our simulations will help clarify this point.

\subsection{Stationary-shape solution}
\label{stationary}

\begin{figure}
\begin{center}
\includegraphics[width=6 cm]{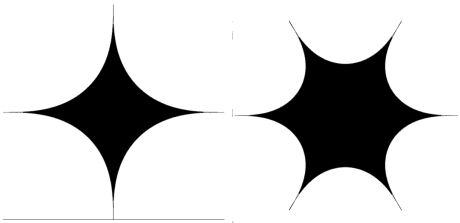}
\caption{ Stationary solutions in a rotating H-S cell with four-
and six-fold symmetry.} \label{station}
\end{center}
\end{figure}

The various scaling behaviors derived so far point at an
asymptotic pattern enclosing a finite area [Eq.
(\ref{scalingarea}); see  Fig. \ref{station}] 
with radial, semi-infinite [Eq.
(\ref{scalingr})] and straight [Eqs. (\ref{scalingr}) and
(\ref{scalingh})] filaments of zero width [Eq. (\ref{scalingh})]
(see Fig. \ref{station}) with a droplet at their tip.

In a stationary state, $\vec u_{\rm in}=\vec u_{\rm out} =0$ by
definition, and the interface shape must balance exactly surface
tension with the centrifugal force in Eq. (\ref{veljump}):
\begin{equation}
\label{zerovorticity}
\sigma\partial_s\kappa =
\Delta\rho\Omega^2 \vec r\cdot\hat s.
\end{equation}

Noting that $\hat s=d\vec r/ds$, where $s$ is arclength along the
interface, we can integrate the above equation once to find
\begin{equation}
\label{intzerovorticity} \kappa(s) = \kappa(s=0) +
\Delta\rho\Omega^2/(2\sigma) \; [r^2(s)-r^2(s=0)].
\end{equation}
We see that the curvature of the interface $\kappa$ increases
monotonically with the distance $r$ to the rotation axis. In
general, the interface will go from convex to concave as $r$ is
increased. The solutions of Eq. (\ref{zerovorticity}) which do
change concavity do not resemble our observed patterns, but have a
petal shape and are unstable \cite{elastica}. We hence turn to
{\em purely convex} stationary solutions of Eq.
(\ref{zerovorticity}), up to the point where $\kappa=0$. These
will represent the asymptotic incoming fingers. To connect them
with the radial, straight filaments, we impose locally radial and
straight ends at an arbitrary distance $r(s=0)$ from the rotation
axis as boundary conditions. Any other connection would cause
$\partial_s\kappa$ in Eq. (\ref{zerovorticity}) to diverge.

For given experimental parameters $\Delta\rho\Omega^2/\sigma$,
this completely determines the solution, an arch covering a
certain angle. This angle is a continuous function of the only
dimensionless ratio we can construct, $\Delta\rho\Omega^2
r(s=0)^3/\sigma$, and hence only for discrete values of this ratio
is the angle a submultiple of $2\pi$. This then allows one to
connect an integer number of equal arches to form an apparently
closed interface, as we show in Fig.\ \ref{station}. This could
not be a stationary solution on its own, since actually closing
the interface would imply that the arches meet in cusps. We
continue each arch end by a straight, semi-infinite radial line,
replacing each cusp with an infinite, straight filament of zero
width. These filaments present no flux inside (finite velocity but
zero width) nor induce any flow outside. They are hence compatible
with stationary arches. Finally, the interface is closed by
placing a droplet of arbitrary size at the tip of each infinite
filament. Although a moving droplet does induce an outside flow
around it, this finite disturbance vanishes near the center for an
infinite filament.

Thus, we have constructed a solution which shape is stationary at
finite distances. Its physical relevance, however, remains to be
checked. The slowing down of incoming fingers derived above and
the flattening of an initially curved filament $h(x)$ [Eqs.
(\ref{scalingr}) and (\ref{scalingh})] are suggestive that such a
solution can indeed be an attractor of the dynamics for initial
conditions with the appropriate n-fold symmetry, and the numerical
simulations in next section and the results of Sec. \ref{lubr}
further support this idea.

A related issue is the selection of the length scale $r(s=0)$ of
the stationary arches. For fixed boundary conditions, only
discrete values of the dimensionless ratio $\Delta\rho\Omega^2
r(s=0)^3/\sigma$ allow equal arches to meet without crossing.
Therefore, for given experimental parameters, the stationary-shape
solutions only exist for discrete values of $r(s=0)$ (note that
unequal arches cannot meet at a {\em single} distance $r(s=0)$
from the center). This sets up a selection problem for the values
of $r(s=0)$.

Dynamically, such a solution will be approached with a certain
number of fingers. This number may not be unambiguously defined
for unequal finger configurations, and it is indeed a dynamic
variable for high viscosity contrasts. Furthermore, the initial
condition is generically nonsymmetric, with some fingers competing
and advancing over others. As a consequence, real experiments and
numerical simulations can evolve towards configurations close to
the above stationary-shape solutions during a transient but will
eventually depart from them. However, for initially equal fingers
as those one can produce in simulations (see next section), the
dynamics conserve the number of fingers regardless of the
viscosity contrast. Hence, only the stationary-shape solution of
length scale $r(s=0)$ with the number of fingers of the initial
condition can be approached. This then unambiguously selects the
scale $r(s=0)$ and the area enclosed by the stationary interface
$a(t\rightarrow \infty)$ (Eq.\ \ref{scalingarea}).
%amount in the problem: the droplet size.

We thus see that the combination of the scaling solutions and the
stationary state derived could specify the asymptotic dynamics of
equal fingers and the ultimate interface shape. As for its
relevance for unequal fingers, we point out two facts: First, we
find reasonable agreement between the exponential rate predicted
by this simple theory and that measured experimentally for the
radial position of the droplets at the tip of the filaments.
Second, the existence of the above symmetrical stationary
solutions could explain the experimental observation of plateaux
in the evolution of the area of inner fluid contained in a
reference circle, as a function of time in Ref. \cite{Alvalow03}.
In fact, in those experiments the interface could successively
visit the neighborhood of stationary states for different numbers
of fingers, and correspondingly slow down the dynamics close to
them until some event (typically a pinch-off) would trigger again
the dynamics.

\section{Numerical tests}
\label{numerics}

We test the basic picture presented above by phase-field
simulations of the full Hele-Shaw dynamics. They are compared to
experimental results and to a numerical solution for the
stationary arches.

\subsection{Method}
\label{method}

First we adimensionalize the problem: lengths will be measured in
units of the initial drop radius $R_0$, and time in units of $t^*$
as given by Eq. (\ref{timescale}). All quantities appearing below
will be dimensionless. The scaling laws in the previous section
can be casted into dimensionless form just by considering all
quantities appearing there dimensionless and redefining
$K=2/(1+A)$ and $K'=1$.

The standard Hele-Shaw dynamics is defined by the
incompressibility condition already used before
($\vec\nabla\cdot\vec u_i = 0$), together with the continuity of
normal velocities across the interface and the tangential velocity
jump [Eq. (\ref{veljump})] as boundary conditions. Thanks to
incompressibility, we can define a stream function $\vec
u=\vec\nabla\times\psi\hat z$ (where $\hat z$ is the direction
perpendicular to the plates), which is continuous across the
interface. We thus obtain the following governing equations
\cite{Tryggvason85,folch00}:
\begin{subequations}
\label{fbp}
\begin{eqnarray}
\label{irrotational}
\nabla^2\psi=0, \\
\label{normalvel}
\partial_s\psi|_{\rm in} =
\partial_s\psi|_{\rm out} =
-v_n \\
\label{vorticity}  {\frac{\partial\psi}{\partial n}}|_{\rm out} -
{\frac{\partial\psi}{\partial n}}|_{\rm in} = \Gamma,
\end{eqnarray}
\end{subequations}
where the first expresses that the flow is irrotational in each
fluid, the second is the continuity of normal velocities, and the
third, their tangential jump.  The magnitude of this jump is the
strength $\Gamma$ of a divergent vorticity peaked on the
interface, $\Gamma\equiv\gamma+A(\partial_n\psi|_{\rm in}+
\partial_n\psi|_{\rm out})$, where $\gamma/2\equiv (B \vec\nabla\kappa
- r \hat r ) \cdot \hat s$ is its local part. We see that the
dynamics are controlled only by two dimensionless parameters: the
viscosity contrast $A$ and the dimensionless surface tension
$B=\sigma/ [\Delta\rho\Omega^2 R_0^3]$ (the ratio of stabilizing
to destabilizing forces). Remarkably, none of the scalings derived
in the previous section depends on $B$.

Simulations were run using a phase field model presented in
\cite{pf1} and extensively tested in \cite{pf2}:
\begin{subequations}
\label{pfm}
\be
\epsilon \frac{\partial\psi}{\partial t} =\nabla^2\psi+A\vec
\nabla \cdot (\theta \vec \nabla \psi)+\frac{1}{\epsilon}
\frac{1}{2\sqrt 2} \gamma(\theta )(1-\theta^2),\label{eq:sf}  \ee
\be \epsilon^2 \frac{\partial \theta}{\partial t} & = &
f(\theta)+\epsilon^2\nabla^2\theta +\epsilon^2 \kappa(\theta )
|\vec \nabla \theta | \nonumber \\& + & \epsilon^2 \hat z \cdot
(\vec \nabla \psi \times \vec \nabla \theta),  \label{eq:pf} \ee
\end{subequations}
where $\theta$ is the phase field, an auxiliary field
distinguishing between the two fluids, $f(\theta )\equiv \theta
(1-\theta^2)$, $\frac{\gamma(\theta)}{2}\equiv \hat
s(\theta)\cdot(B\vec\nabla \kappa(\theta) +\hat y)$ and
$\kappa(\theta)\equiv -\vec\nabla \cdot \hat n(\theta)$, with
$\hat n(\theta)\equiv \frac{\vec\nabla \theta}{|\vec\nabla
\theta|}$ and $\hat s(\theta)\equiv \hat n(\theta) \times \hat z$.

Apart from the physical control parameters $A$ and $B$, the
dynamics in this model also depend on an artificial interface
thickness $\epsilon$ and a relaxation time for the stream function
(which is diffusive, not Laplacian) $\tilde{\epsilon}$. In the
limit $\epsilon, \tilde{\epsilon} \rightarrow 0$,  the dynamics
are strictly those of Eqs. (\ref{irrotational}--\ref{vorticity}).
For finite values of $\epsilon,\tilde{\epsilon}$, an error bound
is guaranteed for any given magnitude by conveniently decreasing
$\epsilon,\tilde{\epsilon}$ \cite{pf2}.

A very interesting feature of this type of model is that both
fluids and the interface between them are treated as bulk. One
consequence of this is that the model is well behaved as two
interfaces break up and reconnect, which enables us to study the
dynamics after pinch-off in a very natural way.
This has been demonstrated very recently \cite{pinch3d}
for the model used here \cite{pf1,pf2}
reformulated in terms of the velocity vector.
Let us add a note of caution: When two
interfaces approach to distances comparable to their thicknesses
$\epsilon$ (more precisely, below $5\epsilon$ from numerical
tests) they attract each other. This triggers the pinch-off. One
could think of this effect as a phenomenological ``pinch rule''.
We nevertheless
do not pretend the phase-field dynamics near pinch-off to
represent the effect of the third dimension nor the eventual
breakdown of a hydrodynamic description. Our approach is to
simulate {\em only} the bidimensional Hele-Shaw dynamics down to
distances of ${\cal O}(\epsilon)$, and continue to simulate them
once the reconnected interfaces have separated a distance of the
same order, very much as in the experiments, where the Hele-Shaw
dynamics can be considered to be accurate before and after the two
interfaces were separated a distance of ${\cal O}(b)$. The
crossover from the Hele-Shaw dynamics to interface interaction
{\it via} their finite thicknesses is very abrupt, enabling us to
clearly separate one from the other.

For reliable quantitative comparison with theory and experiments,
we explicitly check convergence of the time constant $K$ in
$\epsilon,\tilde{\epsilon}$, since $\tilde{\epsilon}$ conveys some
finite diffusion time to the flow, and $\epsilon$ delays the
interface advance with respect to the normal fluid velocities
\cite{pf1}. We thus establish exponential rate values for the
$\epsilon,\tilde{\epsilon}\rightarrow 0$ limit, i.e., for the
Hele-Shaw dynamics, and bounds for $\epsilon,\tilde{\epsilon}$ to
obtain those values in practice. For $\tilde{\epsilon}$, a value
of $\tilde{\epsilon}=0.5$ turns out to suffice, and it is used in
the following unless otherwise stated. Convergence in $\epsilon$
is discussed case by case.

We numerically integrate the above model with an Euler scheme and
centered differences. The time step $dt$ is taken to be close to
the stability limit; $dt=0.2\tilde{\epsilon}dx^2$ unless otherwise
stated, where $dx$ is the mesh size. Convergence of the solution
is also tested changing $dx$ ($dx=\epsilon$ generally suffices as
shown in \cite{pf2}). Our initial condition is some perturbation
of a centered circle of unit radius, i.e., $r=1+\Delta r(\varphi)$
where $r$, $\varphi$ are polar coordinates with respect to the
rotation axis and $\Delta r(\varphi)< 1$. In the next two
subsections, we consider only identical fingers (n-fold symmetry),
$\Delta r(\varphi)=(2\pi/n) q \cos (n\varphi)$, where $q$ is the
amplitude to wavelength ratio of the perturbation, and $n$ is the
(integer) number of fingers. Unless otherwise stated, we use a
dimensionless surface tension $B$ for which $n$ is the most
unstable mode in the linear regime and $q=0.05$.

\subsection{Filament thinning}
\label{thinning}

\begin{figure}
\begin{center}
\includegraphics[width=5.8 cm]{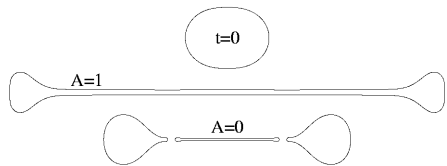}
\caption{Filament thinning: Initial condition (top, $q=0.05$) and
one later interface for $B=0.09$, $\epsilon=0.008$,
$dx=\epsilon/2$ and $A=1$ (middle) or $A=0$ (bottom).}
\label{dumbbell}
\end{center}
\end{figure}

We test here the filament thinning scaling by looking at a
dumbbell-shaped pattern: two droplets connected by a filament
($n=2$, see Fig. \ref{dumbbell}). This case has two advantages: a
width can be unambiguously defined as that at the axis of rotation
(the midpoint between the two droplets); on the other hand, there
are no neighboring filaments to influence its dynamics.

\begin{figure}
\begin{center}
\includegraphics[width=7 cm]{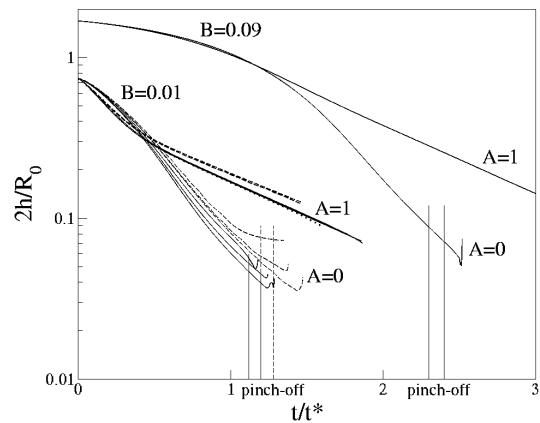}
\caption{Filament thinning: Width at rotation axis (in log scale)
{\it vs.} time. Dashed, solid and dotted line(s): $dx=\epsilon$,
$dx=\epsilon/2$ and $dx=\epsilon/4$, respectively. Upper curves
($B=0.09$): $\epsilon=0.008$ (runs in Fig.\ \ref{dumbbell},
$q=0.05$). Lower curves ($B=0.01$, $q=0.2$): three curves at
$\epsilon=0.02,$ 0.01 and 0.005 for $dx=\epsilon$ and three at
$\epsilon=0.02,$ 0.01, and 0.0067 for $dx=\epsilon/2$ shown for
each value of $A$. For $A=0$, lower curves correspond to lower
values of $\epsilon$. The only $dx=\epsilon/4$ curve is for $A=1$,
$\epsilon=0.02$. The dashed (solid) vertical lines indicate the
interval during which the droplet at the tip pinches off for
$dx=\epsilon$ ($\epsilon/2$). } \label{thinningsimul}
\end{center}
\end{figure}

Figure \ref{dumbbell} displays the patterns thus obtained at the
end of our run ($A=1$) or just after the first pinch-off ($A=0$).
In the last case, new pinch-off events quickly further shorten the
central filament from its ends. We will refer to this phenomenon
as ``pearling''. The two upper curves in Fig. \ref{thinningsimul}
($B=0.09$) show the evolution of the filament width at the
midpoint for the whole run ($A=1$) or up to the relaxation of the
last central segment in a round shape, which makes the width
increase again ($A=0$). The latter occurs when pearling reaches
the rotation axis, shortly after the first pinch-off, indicated by
the vertical lines. We see that the asymptotic decay is grossly
exponential for $A=1$ as predicted, although it is only observed
for roughly a decade. For $A=0$, the filament thinning is strictly
not exponential since the effective exponential rate varies slowly
in time. It should be noticed that it is the flatness, not the
width of the filament, what determines the goodness of the
approximation of last section. One might think that pinch-off
prevents the formation of long and hence straight enough filaments
for $A=0$, but, indeed, just before it the $A=0$ filament is just
as straight as the $A=1$ one at the same time; the $A=1$ filament
already exhibits a clear exponential behavior, whereas the $A=0$
one does not.

A closer look reveals that the thinning rate is well relaxed to
its asymptotic value for the last third of the run for $A=1$, with
a value of $K=0.98$ consistent with the predicted $K$, but not for
$A=0$ because pearling reaches the center when the rate begins to
approach (from above) a value of $K=2$. No disturbance is observed
from the previous pinch-off events. Runs for $A=0.5$ and $A=0.8$
(not shown) are also in reasonable agreement with the law
$K=2/(1+A)$.

One might blame the high dimensionless surface tension $B=0.09$
used or the finite interface thickness $\epsilon$ or relaxation
time $\tilde{\epsilon}$ in the model for either the slow
relaxation of the exponential rate or the pinch-off itself in the
low-$A$ case, since $B$ enters the problem as long as the filament
is not perfectly straight. The lower curves in Fig.
\ref{thinningsimul} for a lower dimensionless surface tension,
$B=0.01$, and at different values of $\epsilon$ and $dx/\epsilon$
rule out both possibilities. Using a more deformed circle
($q=0.2$), to avoid the appearance of other modes which are
linearly unstable, we can confirm the results obtained for
$B=0.09$. For high viscosity contrast, the good decay even for the
coarsest mesh and interface profile allows to establish the value
of the rate $K$ unambiguously at $K=0.99-1.00$ by studying the
convergence in $\epsilon$ and $dx$. For low viscosity contrast
runs ($A=0$), the values of the rates again seem to relax towards
the end of the runs to a value compatible with the theoretical
value $K=2$ although they are again badly defined.

Pinch-off is systematically observed for low ($A=0$, 0.5), but not
high ($A=0.8$, 1) viscosity contrasts. For the latter, a numerical
instability due to noise in the (relatively) inviscid fluid often
sets in abruptly at late times, especially for the coarser
discretization $dx=\epsilon$. Because it is so abrupt, it is easy
to separate from valid data. For the former, the pinch-off itself
cannot be considered a spurious effect due to a finite interface
thickness, since the pinch-off time increases only marginally when
decreasing the interface thickness $\epsilon$, and in any case
stays bounded by the vertical lines in Fig. \ref{thinningsimul}.
Only for the coarsest interface profile can we detect that the
center of the filament anomalously ``anticipates'' the first
pinch-off at its tip; for lower values of $\epsilon$ it actually
continues to narrow unaffected after the first pinch-off and until
the subsequent pearling reaches it. Interestingly, when decreasing
the mesh size $dx$ from dashed to solid lines, pinch-off is
appreciably accelerated, so that it must also remain in the
continuum limit.

We conclude that the theoretical exponential thinning and the
values of the exponential rates derived in last section are a good
approximation to the observed behavior, especially for high
viscosity contrasts. Deviations do exist and are intrinsically
much stronger for low viscosity contrasts, but, because they are
not less apparent for a surface tension almost one order of
magnitude lower, they cannot be due to it.

\subsection{Stationary-shape solution}
\label{steadysimul}

We now increase the number of fingers arising from the linear
regime to $n=4$, so that the incoming fingers enclose a finite
area. This enables us to test the approach to the stationary-shape
composite solution derived in Sec. \ref{stationary}.

\begin{figure}
\begin{center}
\includegraphics[width=7.5 cm]{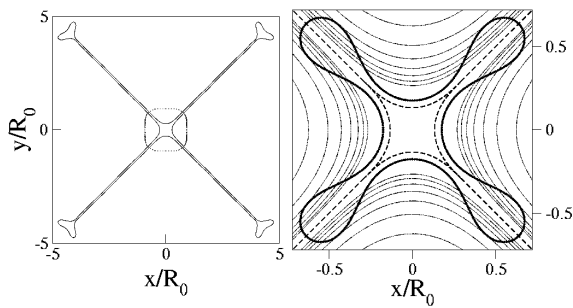}
\caption{Approach to the stationary shape for $B=0.0213$ and
$\epsilon=0.01$. Left: initial (dotted) and latest (solid line)
interfaces for $A=1$, $dx=0.5\epsilon$ and $\tilde{\epsilon}=0.5$.
Right: Blow-up of central region with intermediate interfaces
(solid lines) at constant time intervals. Thicker line: receding
interface (after pearling) for $A=0$, $dx=\epsilon$ and
$\tilde{\epsilon}=1$. Dashed line: numerical solution of Eq.
(\ref{intzerovorticity}) for the predicted stationary shape.
}\label{steadyfig}
\end{center}
\end{figure}

Figure \ref{steadyfig} (left) shows the initial condition (dotted
line) and the last interface (solid line) of our most refined
($\epsilon=0.01$, $dx=0.5\epsilon$ and $\tilde{\epsilon}=0.5$) run
for $A=1$, just before spurious pinch-off occurs. The latter is
unavoidable because the filament width reaches 0.05--0.07, at the
edge of diffuse-interface attraction ($\sim 5\epsilon$). $A=0$
runs pinch much faster, presumably due to a physical phenomenon
and not to the finite interface thickness, since here, as well as
for $n=2$, this is not cured by decreasing $\epsilon$. Due to this
different behavior, $A=0$ runs do not lead to the star-like
pattern shown here, but, after a first pinch-off event at moderate
radii, they quickly emit several droplets while the central drop
recedes.

The blowup in Fig. \ref{steadyfig} (right) shows the last
interfaces for $A=1$ at constant time intervals (thinner solid
lines) compared to a numerical solution of Eq.
(\ref{intzerovorticity}) for the composite stationary pattern
derived in Sec. \ref{stationary} (dashed line). The latter seems a
plausible asymptotic solution in the sense that it has not been
overcome by the incoming fingers of our run, although visually
these seem to asymptotically approach less curved arches. For a
quantitative estimate, we fit a slowing down of the type of Eq.
(\ref{scalingarea}) [i.e., we fit $a(t\to\infty)$,
$a(t=0)-a(t\to\infty)$ and $K$] both to the time evolution of the
area enclosed by a reference circle of various radii $R_{\rm
ref}$, and to that of the radial coordinate of the incoming finger
tips. We find $K=1.06$ for the area enclosed by a circle of radius
$R_{\rm ref}=1.25$, where the filament begins to look straight,
and $K=0.9-1$ for the incoming finger tip. Both are in good
agreement with the theoretical value $K=1$. The area enclosed by
smaller circles decays with smaller exponents (0.93 for a circle
of radius $R_{\rm ref}=1$ and 0.81 for $R_{\rm ref}=0.75$), but a
worse agreement is understandable, since the filaments are less
straight at the latest times we can explore.

Regarding the tip of the incoming fingers and the total enclosed
area in the center, the numerical solution of Eq.
(\ref{intzerovorticity}) predicts $r_{min}=0.133$ and
$a(t\to\infty) = 0.0937$ respectively. The numerical simulations
here presented show that the filament still contains a significant
amount of inner fluid, yielding an asymptotic enclosed area which
decreases with the radius of the reference circle
($a(t\to\infty)=0.36$ for $R_{\rm ref}=1.25$, $a(t\to\infty)=0.27$
for $R_{\rm ref}=1$ and $a(t\to\infty)=0.20$ for $R_{\rm
ref}=0.75$). As for the asymptotic tip position of the incoming
fingers, we find $r_{min}=0.2$, with little sensitivity to the
time window used. We thus seem to find that the tips and the area
enclosed reach roughly half of those predicted by Eq.
(\ref{intzerovorticity}). This mismatch is most probably due to
the fact that we do not yet probe close enough to the
stationary-shape solution. This is indeed the case regarding the
filaments since, as explained above, they have not reached its
zero-width asymptotic state because the final area is not
independent of the reference circle radius (as it should be if
$R_{\rm ref}<r(s=0)$).

However, the evolution of the interface for $A=0$ after pinch-off
does hint that this exact solutions have an important effect on
the dynamics. After the pearling process of the filaments for
$A=0$, the incoming fingers continue to approach the rotation axis
and the droplets left at the tip of the filaments recede. When
they reach the central region roughly delimited by the ends of the
arches of our predicted stationary-shape solution, surface tension
finally stops the advance of the incoming fingers. The whole drop
left at the center relaxes towards a circular shape. The thicker
solid line in Fig. \ref{steadyfig} (right) corresponds to such a
central shape at the time when the incoming fingers stop before
the relaxation to a circular shape. Although the shape displayed
is not an asymptotic one, the remarkable resemblance of the tip of
the incoming fingers and the steady-state asymptotic arches
(dashed lines) and the fact that incoming fingers precisely stop
(their velocity crosses zero) when approaching this shape is
suggestive that our steady-shape solution (or parts of the
solutions found in \cite{elastica} very similar to it) is indeed
the attractor for sets of initial conditions with n-fold symmetry.

\subsection{Unequal finger dynamics}
\label{unequal}

Finally, we address the more realistic case of unequal fingers.
The goal here is no longer to test our theoretical results, but to
directly compare with experimental patterns. We thus
check whether the differences between low and high viscosity
contrasts $A$ observed in experiments, including systematic or not
pinch-off events, can be explained by the Hele-Shaw equations.

In order to compare with the patterns shown in Fig. \ref{morfexp},
which have 15--16 filaments for $A=1$ [Figs. \ref{morfexp}(a,b)] and
20-21 for low $A\sim 0.45$ [Figs. \ref{morfexp}(c,d)],
we use a dimensionless surface tension of $B=1.03\times10^{-3}$,
which corresponds to a most unstable mode $n=18$.
The idea is to use exactly the same
initial condition and physical and computational
parameters for both low and high viscosity contrast $A$, except for
$A$ itself. Thus, any difference we observe can only be due to
the different values of $A$.

We adopt the less refined choices
$dx=\epsilon$ and $\tilde{\epsilon}=1$ with an interface thickness
down to $\epsilon=0.005$, in order to avoid possible spurious
pinch-off events. As already observed for the dumbbell-shaped
pattern in Sec. \ref{thinning}, we find that, whenever pinch-off
occurs, a lower rate of $dx/\epsilon$ does not prevent it, but
even accelerates it.

As for the initial condition, if we start with a perturbation
where sine and cosine modes have uniformly or Gaussian distributed
random amplitudes up to a certain cut-off (here $n=40$), we cannot
reproduce the strong asymmetry of the experimental patterns for
high viscosity contrast $A=1$ [Figs. \ref{morfexp}(a,b)]. We
traced this back to the fact that, in the
experiments, the deviations from a circle first become visible in
particular spots, and not uniformly on the interface. For the
experiment of Figs. \ref{morfexp}(a,b), this happens roughly on
three spots. We mimic this by multiplying our random initial
condition by an exponential envelope peaked at three uniformly
distributed random angles, keeping $\Delta r(\varphi)\leq 0.004$.
We do not attempt to truly copy the experimental initial
condition, but to use a somehow ``statistically'' similar one.

If some details look alike, these will then
indicate some general common trends representative of the
dynamics, as opposed to a particular initial condition.

\begin{figure}
\begin{center}
\includegraphics[width=0.445\textwidth]{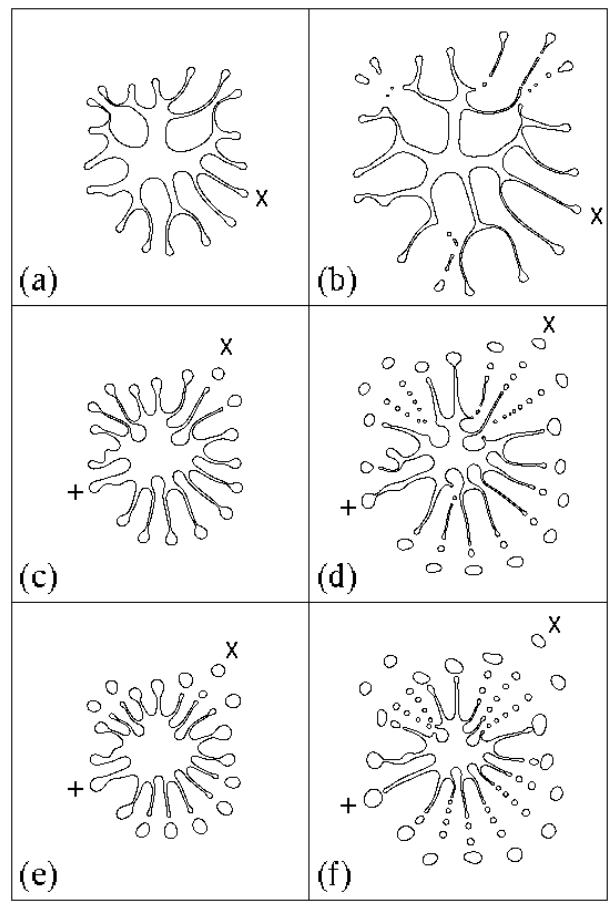}
\caption{Pattern evolution for a random initial condition (see
text) and $B=1.03\times10^{-3}$, $\epsilon=0.005$, $dx=\epsilon$,
$\tilde{\epsilon}=1$ and $A=1$ (a,b), $A=0.5$ (c,d) or $A=0$
(e,f). Snapshots ($3R_0\times3R_0$) shown at $t=1.32$ (left) and
$t=1.74$ (right). }\label{patternsimul}
\end{center}
\end{figure}

This initial asymmetry in the deviations from a circle is largely
preserved throughout the whole evolution
for high [$A=1$; Figs. \ref{patternsimul}(a,b)], but
not low viscosity contrast [$A=0$, $0.5$ runs;
Figs. \ref{patternsimul}(c,d) and Figs. \ref{patternsimul}(e,f),
respectively]. We hence conclude that both an asymmetric initial
condition and a high viscosity contrast are necessary to obtain
asymmetric patterns.
This effect of the viscosity contrast is due to the
incoming finger competition for high $A$, although it is less apparent
here than in the channel geometry, due to the geometric decrease of
available room fingers experience.

Figure \ref{patternsimul} is indeed the simulation analogue of
Fig. \ref{morfexp}. Experimental patterns in Fig. \ref{morfexp}
were rotated at will to make the ``statistical'' resemblance with
the simulations more apparent; simulations are presented with the
vertical and horizontal directions parallel to the computational
grid, although lattice anisotropy is not appreciable here.

Patterns are shown at two different times:
$t=1.32$ [Figs. \ref{patternsimul}(a,c,e)] and
$t=1.74$ [Figs. \ref{patternsimul}(b,d,f)],
for all viscosity contrasts.
Rather than comparing these simulations with
experiments at equal dimensionless times,
we present all the patterns when their envelope has
roughly attained twice
[Figs. \ref{morfexp}(a,c) and \ref{patternsimul}(a,c,e)] and three
times [Figs. \ref{morfexp}(b,d) and \ref{patternsimul}(b,d,f)]
the initial circle radius. Direct comparison of the time scales is left for
Sec. \ref{sectimescale}.

Let us first compare patterns at the earlier stage [Fig.
\ref{morfexp}(a) with Fig. \ref{patternsimul}(a), and Fig.
\ref{morfexp}(b) with Fig. \ref{patternsimul}(b)]. The similarity
between experiments and simulations is remarkable, especially if
one takes into account that the initial conditions were only
``statistically'' similar (see above). Furthermore, the common
features (overall morphology, filament accomplished or frustrated
branching, droplet size, the already mentioned incoming fingers)
are well reproducible in other simulation runs and experiments.
This all indicates that the excellent agreement in the typical
morphologies between experiments and simulations is not
fortuitous. This is noteworthy, since a change in the wetting
conditions for the high viscosity contrast experiments, for
instance, completely changes the observed morphology
\cite{Alvahigh03}.

>From this earlier stage it would seem that pinch-off is not significantly
more present for the lower ($A=0.5$) than higher ($A=1$) viscosity
contrasts. However, simulation offers us the possibility to go to
the ideal limit $A=0$, in which both fluids have strictly equal
viscosities. Fig. \ref{patternsimul}(e) shows that most droplets have
pinched-off by the same time for $A=0$. Comparing
Figs. \ref{patternsimul}(a,c,e) between them, it is clear that pinching
arises as the viscosity contrast is decreased.

Another possibility is to go to later times. The experiments
still show no pinch-off for $A=1$ [Fig. \ref{morfexp}(b)], but
most filaments have emitted at least one droplet for $A\simeq 0.5$
[Fig. \ref{morfexp}(d)]. The simulations would seem less conclusive,
since they display some pinch-off events for $A=1$
[Fig. \ref{patternsimul}(b)]. However, these are most likely to be
spurious: They occur for very narrow filaments, whose width is
comparable to the interface thickness $\epsilon$,
and we know from earlier work \cite{pf1,pf2}
that overlapping diffuse interfaces attract each other.
Most importantly, pinching is inhibited as $\epsilon$ is decreased
(the same run with a larger value of $\epsilon$ displays more
pinching).

Simulations for low viscosity contrasts
[$A=0.5$, Fig. \ref{patternsimul}(d) and
$A=0$, Fig. \ref{patternsimul}(f)], show much more pinching
(the lower $A$, the more). The first pinch-off at the tip of a filament
does not significantly change with $\epsilon$,
very much like we observed for the dumbbell-shaped pattern in Sec.
\ref{thinning}. Some pinch-off events disappear,
while other appear; some are reentrant in $\epsilon$
(they disappear for a middle value of $\epsilon$
but reappear when $\epsilon$ is further decreased), .
We hence conclude that these first pinch-off events are {\em not}
spurious. In any case, longer and thinner filaments are achieved as
$\epsilon$ is decreased.

Pearling (successive pinch-off events shortening a filament from
its outer end) is {\em increased} when decreasing $\epsilon$, even
though it is less present in the experiments [Fig.
\ref{morfexp}(d)]. It is interesting to note the absence of
pinching at the base of the filaments, near the rotation axis, for
$A=0$ [Fig \ref{patternsimul}(f)], as opposite to the $A=0.5$ and
$A=1$ runs [Figs \ref{patternsimul}(b,d), respectively]. Since we
saw pinching to be spurious for the $A=1$ run, this suggests that
such events at the base of a filament are also spurious.
Furthermore, they are not observed in the experiments.

We conclude that simulations reproduce very accurately the
experimental morphologies before pinching, which strongly supports
the idea that the dynamics can be considered to be the standard
Hele-Shaw for that purpose. They also indicate that these
bidimensional dynamics do lead to finite-time pinching for low,
but not high viscosity contrasts. The crossover occurs somewhere
between $A=0.5$ and $A=1$.

\subsection{Filament stretching and droplet dynamics}
\label{stretchingsimul}

We have postponed the study of the filament stretching
and droplet scalings to present it in a unified way, for all the
cases ($n=2$, 4 and 18 filaments) considered above.
Let us first summarize what we can expect from the
theory in Sec. \ref{basics}: Droplets at the tip of a filament should
run away roughly exponentially in time,
with a rate comprised between $K'=1$ (for an isolated droplet)
and $K=2/(1+A)$ (the filament thinning rate).
The lower bound $K'$ corresponds to the idealized case where the
filament does not perturb the droplet shape nor the flow created by it,
so that the droplet still obeys the exact solution
for an isolated and circular one. This assumption is more realistic
for high viscosity contrasts, since we have seen that in the $A\to 1$
limit the homogeneous flow inside an isolated circular droplet
holds for any closed interface, regardless of its shape and whether
it is isolated or not.
Note, however, that a droplet at the tip of a filament is not
completely closed, so that a deviation from $K'$ is still possible.

The upper bound $K$ is based on another idealization, where no
fluid is fed into the droplet, so that all the fluid expelled by
the filament as it narrows is used to extend it. Therefore, the
radial coordinate of the neck of the filament just before the
droplet scales with the same exponent $K$ as the filament width.
If the shape of the droplet is stationary, the radial coordinate
of the droplet center of mass is then just that of the neck of the
filament plus a constant.

The purpose of this section is to determine which is the actual
evolution of the radial coordinate of the droplets center of mass.
Our experiments have not clearly answered this question: Droplets
are seen to (i) run away with roughly the same rate whether
connected to a filament or isolated, but (ii) this observed rate
seems to be $K$ (filament thinning) rather than $K'$ (isolated
droplet). The problem is that (ii) is inconsistent with the {\em
exact} isolated droplet solution.

\begin{figure}
\begin{center}
\includegraphics[width=7 cm]{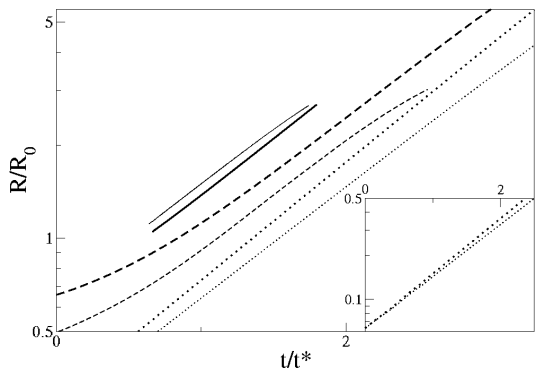}
\caption{Radial position (in log scale) of the center of mass of
various droplets vs. time, for $A=1$ (bold) and $A=0$ (thinner
curves). Solid, dashed and dotted lines correspond to the droplets
indicated with an ``x'' in Figs. \ref{patternsimul}(a,b,e,f), those
in Fig. \ref{dumbbell}, and the isolated droplet in Fig.
\ref{stream} bottom-right, respectively.  Curves are shown till
the end of the respective runs, except for the isolated droplet,
which is shown only until it begins to ``feel'' the finite
simulation box. Inset: Linear regime of the isolated droplet,
continued in the main plot after translating the (dotted) curves
to earlier times for comparison with the others. The inset
preserves slopes with respect to the main plot. }
\label{dropletscaling}
\end{center}
\end{figure}

Fig. \ref{dropletscaling} shows the evolution of the radial
coordinate of a droplet center of mass for the dumbbell-shaped
pattern in Fig. \ref{dumbbell} ($n=2$, dashed lines) and some of
the furthest droplets
in Fig. \ref{patternsimul} ($n=18$, solid lines), compared
to that of an isolated circular droplet ($n=1$, dotted lines).
Bold (thinner) lines correspond to $A=1$ (0).
Curves are shown till the end of the respective runs, except for
the isolated droplet, which is shown only until it begins to ``feel''
the finite simulation box. Note that for $n\ge 2$ curves make
little sense for $R<R_0$, and this part has been removed for the
$n=18$ runs, since the code actually detected a different droplet
at those early stages. More details on how the curves were obtained
can be found below. Although the range of $R/R_0$ monitored
decreases with increasing $n$, the final aspect ratio of the pattern
(droplet radius to length of filament for $n\ge 2$, or droplet
radius to radial coordinate of center of mass of the isolated droplet)
attained is similar.

Clearly, all droplets scale with roughly the same rate $m$,
although runs with $A=1$ show a good linearity, whereas those for
$A=0$ do not. More precisely, for $A=1$ (0), one measures (in the
straightest segment of the curves) $m=0.8$ (0.7) for $n=2$,
$m=0.82$ (0.75) for $n=18$, and $m=0.88$ (0.83) for $n=1$. Runs
for $n=4$, 6, 8, 11 and 12 fingers, give similar results, also for
intermediate values of the viscosity contrast ($A=0.5$, 0.8).

Since all droplets scale with a same rate, roughly independent
of $A$ and close to $K'=1$,
it is clear that the droplet scaling is dominated by that of
an isolated droplet. This rate is found to be insensitive to pinch-off,
and it also holds for the experimental isolated droplet
in Fig. \ref{dropletscaling} (dotted lines), which
further supports this idea,
and confirms this part of our experimental conclusions (i).
However, the fact that droplets scale roughly with $K'=1$ for
any viscosity contrast $A$ in our simulations ($m\sim 0.8$) and
certainly not with $K=2/(1+A)$ confirms that there is some problem
with the statement (ii) that the experimental
growth rate be $K=2/(1+A)$, as already expected from
the exact isolated droplet solution.
We address this issue
in the following section.

The mismatch
between the droplet scaling ($m\sim 0.8$) and the filament thinning
$K=2/(1+A)$ accounts for the fluid fed into the droplet. This is
relatively small for $A=1$ ($K=1$), but not for
$A=0$ ($K=2$). This explains the experimental and numerical
observation that droplets grow
faster and larger as $A$ is decreased.
The bad linearity of the $A=0$ runs is also probably due this large
mismatch, although we recall that the filament thinning itself
displays a less constant exponent for $A=0$ than $A=1$.

The center of mass of the droplets for $n=2$, 18 in Fig.
\ref{dropletscaling} was computed subtracting the final droplet
radius to the droplet tip position, which is not rigorous, because
the droplet radius changes during the evolution. However, if one
subtracts, say, the final distance from the droplet tip to the
neck of the filament before the droplet for comparison, rates
increase by 0.1 only (not shown). We have also extracted the neck
position (not shown) of the same droplets (marked with an ``x'')
in Fig. \ref{patternsimul}, and measured an exponent up to $m=0.9$
(for $A=0.5$). We also confirmed that a droplet which pinches only
at the end of the run (marked with ``+'') scales with the same
rate than one of the first to pinch (marked with ``x'').

\begin{figure}
\begin{center}
\includegraphics[width=7 cm]{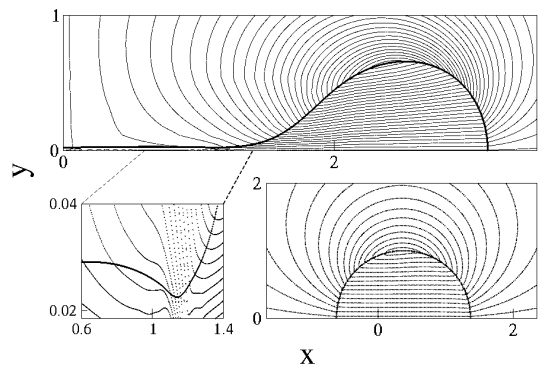}
\caption{Stream lines induced for $A=0$ by a dumbbell-shaped
pattern as that in Fig. \ref{dumbbell} (top; a quarter of the
interface shown, $B=0.09$, $\epsilon=0.0067$, $dx=\epsilon/2$,
$\tilde{\epsilon}=0.5$) or by an off-center circular droplet
(bottom right; a half shown, $B=0.32$, $\epsilon=0.02$,
$dx=\epsilon$, $\tilde{\epsilon}=0.1$,
$dt=0.25\tilde{\epsilon}dx^2$). Bottom left: Blow-up ($y$ enlarged
30$\times$ more than $x$ coordinate) of the forming neck in the
top pattern. Interface (thicker lines) and contour plot of the
stream function (thinner lines) at equally spaced levels (different in
each graph); $x=y=0$ is the rotation axis. }\label{stream}
\end{center}
\end{figure}

In the end, the main effect of the filament on the scaling of the
center of mass of the droplet seems to come from the injection of
inner fluid into it, rather than perturbing the droplet
environment. This should mean that the flow of the inner fluid is
much more affected than that of the outer by the presence of the
filament. The latter is illustrated for the most sensitive case
($A=0$) in Fig. \ref{stream}, where stream lines are superimposed
to the interface for a quarter of a dumbbell-shaped pattern (top),
as that in Fig. \ref{dumbbell}, and for half an isolated droplet
(bottom right): The most remarkable difference is the higher
density of stream lines (higher velocity) at the entrance of the
droplet and {\em inside} it, which breaks the flow uniformity
within the droplet; the outer flow is qualitatively similar near
the droplet (note that stream lines are not shown at the same
levels for a maximum resolution of the flow details in both cases,
so that the modulus of the velocity cannot be compared).

When a neck is formed (blowup, bottom left, in Fig. \ref{stream}),
the flow is affected also outside the droplet, but mostly
near the neck. Because pinching then also occurs in a time scale sensibly
shorter than that of the filament stretching, the effect is not perceived
in the latter. The $A=0$ droplets tracked in Fig. \ref{dropletscaling}, for
instance, pinch during $t\in (0.9,0.96)$ for $n=18$ and
$t\in (2.3,2.4)$ for $n=2$, but they curve down well after or
before it, respectively.

One could also wonder why the exponents measured for the
isolated droplet run are not strictly $m=K'=1$, but this is certainly
a finite-interface-thickness correction due to the less refined value
used ($\epsilon=0.02$). Our purpose here was merely to check
that the exact solution for the isolated droplet has some finite
basin of attraction, as we find it to be the case:
The droplet stays circular
and runs away exponentially with a very steady rate
from the linear ($q=0.005$, inset of Fig. \ref{dropletscaling})
to deep into the nonlinear (main plot) regimes.
The droplet depicted in Fig. \ref{stream} bottom right
corresponds to the end of the linear (and beginning of the nonlinear)
regime. The type of flow predicted by the exact solution
[Eqs. (\ref{dropletflow})], is apparent here.

\section{Time scale}
\label{sectimescale}

In Sec. \ref{stretchingsimul} we saw that the statement that the
experimental growth rate be $K=2/(1+A)$ and not $K'=1$ is
inconsistent not only with the exact isolated droplet solution,
but also with numerical simulations of the Hele-Shaw equations.
This statement relies only on the observed $A=0.45$ rates, since
for $A=1$ $K=K'=1$ (in good agreement with the measured values).
This indicates that our experiments for low $A$ do not follow the
standard Hele-Shaw dynamics.

A three-dimensional study of the proper effective boundary
conditions to apply on an idealized two-dimensional interface, as
performed for air displacing a liquid \cite{Park}, is lacking for
the case of two liquids. This low-$A$ case might be different,
since a second viscosity is involved when a layer of vaseline oil
(outer fluid) wets the plates also inside the silicone oil (inner
fluid) domain. Note that, for $A=1$, such a wetting layer (in this
case, of silicone oil) is to be found in the air. However,
different boundary conditions should in general not produce
experimental patterns so similar to the outcome of simulations
using standard boundary conditions as those in Sec. \ref{unequal}.
Especially, they should not necessarily preserve the pinch-off for
low $A$.

The inferred dimensionless rates for the experimental droplet
runaway rely on the measured rates and on the time scale $t^*$. A
modification of the latter can hence already account for the
observed discrepancy between theory and simulations. To further
realize this, it is instructive to compare again experimental
(Fig. \ref{morfexp}) and simulation (Fig. \ref{patternsimul})
patterns: The similarity in the morphologies is obvious, but time
scales are not straightforward to compare, since the degree of
deviation from a perfect circle in the experimental initial
condition is unknown. This results in an effective ``latency
time'', during which no significant departure from a circle is
observed.

However, we can compare the time elapsed for the pattern envelope
to grow from a radius of $2R_0$ [Figs. \ref{morfexp}(a,c)
and \ref{patternsimul}(a,c,e)] to
$3R_0$ [Figs. \ref{morfexp}(b,d)
and \ref{patternsimul}(b,d,f)].
This is $\Delta t =$6s(36s) for A=1(0.45) in the experiments.
With the measured values of the physical parameters,
$t^*$ in Eq. \ref{timescale} is $t^*=15.5$s(134s),
which results in a dimensionless time interval
$\Delta t/t^*=0.39(0.27)$ for A=1(0.45).
This is to be compared with $\Delta t/t^*=0.42\;\;\forall A$
in the simulations.
The mismatch between experiments and simulations is not
significant (taking into account the viscosity uncertainty and the
subjectivity in the snapshots to compare)
for $A=1$, but it is for $A=0.45$.
This is just a (statistical) confirmation of the droplet
scaling, since it is this scaling what determines the growth of
the pattern envelope, but makes it clear that the main difference
between experiments and simulations is
the time scale for $A=0.45$.

A simple explanation is that the vaseline wetting layer has a more
obvious effect than modifying the effective
two-dimensional boundary conditions on the interface:
it changes the expression for Darcy's law [Eq. (\ref{darcy})].
The two-dimensional velocity in Darcy's law results from
an average over the cell gap of the three-dimensional velocity field.
We simply point out that one should average over the fraction of the
gap actually filled with
silicone oil when computing the averaged silicone oil velocity.
Assuming stick boundary conditions for the vaseline on the glass
plates, and no-slip between vaseline and silicone, we average
the silicone velocity in a region closer to its maximum (at the midplane
between the two glass plates) than in the absence of the vaseline
wetting layer. This results in a higher velocity, further enhanced by
the fact that vaseline is less viscous. More precisely, we find
\be
\label{newdarcy}
\vec u_{\rm in}=-\frac{b^2 [1+\Delta b^2]}{12\mu_{\rm in}}
( \vec\nabla p_i - \rho_i \Omega^2 r \hat r ),
\ee
where
\be
\Delta b^2=\frac{2w}{b}
+\left (8+6\frac{\mu_{\rm in}}{\mu_{\rm out}}\right )
\left (\frac{w}{b}\right )^2,
\ee
with $w$ the thickness of the vaseline wetting layer at each glass plate.
The effect of this on the time
scale of the dynamics $t^*$ in Eq. (\ref{timescale}) is to replace
$\mu_{\rm in}$ there by $\mu_{\rm in}/[1+\Delta b^2]$
($\vec u_{\rm out}$ stays unchanged).
For a wetting layer of thickness $w$ 10\% of the total cell gap $b$,
our $A=0.45$ time scale should decrease by a factor 0.78. This
resets the rate at which droplets run away to $m=0.97$--1.17, in
reasonable agreement with theory and simulations. We have not
measured the thickness of the wetting layer $w$, and possibly other
effects like the effective two-dimensional boundary conditions on
the interface might play a role, but this gives a plausible explanation
of our observations.

\section{Lubrication theory}
\label{lubr}

The purpose of this section is to derive the time evolution of a
gently curved filament in a more systematic way. In particular, we
would like to account for the striking difference between the
filament thinning of high and low viscosity contrast dynamics.

To do so, we perform a lubrication approximation for the
interface. This assumes the interface height $h(x,t)$ in Fig.
\ref{sketch}  to vary in a scale $\Delta h$ much smaller than the scale
of horizontal variations $\ell$, and expands formally all
quantities in powers of $\varepsilon=\Delta h/\ell$. We then find an
evolution equation for $h$ up to first order in $\varepsilon$. The
idea is to see whether the viscosity contrast can affect the
relative importance of the various terms and to what order it
enters.

We begin by rewriting the free boundary problem of Eqs.
(\ref{fbp}) into an exact evolution equation for $h$: Using the
vortex-sheet formalism of Refs. \cite{Tryggvason85,Alvarez01}, we
have
\be
\partial_t h = U_{\hat{y}} -  U_{\hat{x}} h_{x}, \label{hdet}\ee
where $U(x,t)=U_{\hat{x}}\hat x +U_{\hat{y}}\hat y $ is the mean
fluid velocity at the interface, which we assume to vary only in
the $x$ direction:
\be
\begin{array}{ll}
\vec{U}(x,t) = & \frac{1}{2\pi} P \int_{-\infty}^{+\infty} \frac{(
h(x^{\prime}) -  h(x) , x - x^{\prime}  ) }{ (x - x^{\prime})^{2}
+ ( h(x^{\prime})-h(x) )^{2} } \Upsilon (x^{\prime}) dx^{\prime}
\\
\\ & -  \frac{1}{2\pi} P \int_{-\infty}^{+\infty} \frac{(
-h(x^{\prime}) -  h(x) , x - x^{\prime}  ) }{ (x - x^{\prime})^{2}
+ ( h(x^{\prime})+h(x) )^{2} } \Upsilon(x^{\prime}) dx^{\prime},
\label{tot1}
\end{array}
\ee where $P$ is the Cauchy principal part and $\Upsilon=\Gamma
(1+h_x^2)$ ($\gamma$defined in Sec. \ref{method}) becomes
\be
\Upsilon/2=B\partial_x\kappa - A \vec{U}\cdot (\hat x + \hat
y\partial_x h) + x+h\partial_x h. \ee

We now scale $h$ with $\Delta h$, $x$ with $\ell$, and the interface
velocity $\vec{U}$ with $V_0=(B/\ell^2)+ \ell$. We expand any
quantity $Q$ as $Q=Q^{(0)}+\epsilon Q^{(1)} +\epsilon^2 Q^{(2)}$,
so that the evolution equation for $h$ up to $O(\varepsilon)$
becomes
\be
\partial_t h= U_{\hat{y}}^{(1)} - \partial_x h U_{\hat{x}}^{(0)} + \varepsilon
(U_{\hat{y}}^{(2)}- \partial_x h  U_{\hat{x}}^{(1)}),
\label{silly}\ee where we have anticipated that
$U_{\hat{y}}^{(0)}=0$. $\vec{U}$ and $\Upsilon$ are indeed
expanded along the same lines as in Refs.
\cite{Alvarez01,Goldstein98}, to further find
\be
U_{\hat{x}}^{(0)}=\frac{1}{2} \Upsilon ^{(0)} \qquad
U_{\hat{x}}^{(1)}=\frac{1}{2} \Upsilon ^{(1)}+  H[\partial_x(h
\Upsilon ^{(0)})]\ee
\be
U_{\hat{y}}^{(1)}=-\frac{1}{2}[\partial_x(h \Upsilon ^{(0)})
+h\partial_x\Upsilon ^{(0)}] \ee \be
U_{\hat{y}}^{(2)}=-\frac{1}{2}[\partial_x (h \Upsilon ^{(1)})
+h\partial_x\Upsilon ^{(1)}]-h H[\partial^2_x (h \Upsilon
^{(0)})], \ee
\be
\Upsilon ^{(0)}=\frac{2 L}{(1+A)V_o} \eta(x) x \ee \be \Upsilon
^{(1)} =\frac{2B}{(1+A)V_o L^2}
\partial^3_x h -\frac{2A}{1+A} H[\partial_x (h \Upsilon ^{(0)})], \ee
where $H(x)=\pi^{-1} P\int_{-\infty}^{+\infty} f(x')dx'/(x-x')$ is
the Hilbert transform of $f(x)$, and $\eta(x)$ is an arbitrary
cut-off function in the centrifugal force to render $H$ finite,
which is 1 up to a certain distance $x_F$ and then decreases to
zero in an arbitrary way. We pursue the calculation and check
whether the result is independent of the shape of $\eta$. This
cut-off is necessary, as opposed to the channel case, because the
zero order of the vorticity in a rotating cell is neither zero nor
a constant but, $\Upsilon ^{(0)} \propto x$. The other major
difference between the weakly nonlinear expansion performed in
\cite{Alvarez01} and this lubrication approximation is the
appearance of the second integral in (\ref{tot1}). Its expansion
is not trivial as pointed out in \cite{Goldstein98}; there it was
solved using limiting procedures, the Plemelj formula, and delta
function representations.

Substituting these results into Eq.(\ref{silly}), undoing only the
scaling of the velocity, and dropping the cut-off function
$\eta(x)$, which is not necessary for finite or asymptotically
straight infinite filaments,
we finally obtain

\be
\begin{array}{ll}
\frac{1+A}{2}\partial_t h= & - \partial_x(x h) - \varepsilon
B\partial_x(h\partial^3_x h) \\ \\&
%-\frac{1-A}{1+A}
-\varepsilon \frac{(1-A)}{1+A} \partial_x\left
(hH[\partial_x(xh)]\right ).
\end{array} \label{expandedevolution} \ee

We see that this is a simple continuity equation for $h$, namely
that already used in Sec. \ref{scalings}, $\partial_t
h=-\partial_x(hv_{\hat{x}})$ [Eq. (\ref{hcontinuity})], but now
with an inner fluid velocity along the $x$ direction
\be
v_{\hat{x}}=\frac{2}{1+A}\left ( x + \varepsilon B\partial^3_x h +
\varepsilon \frac{1-A}{1+A} H[\partial_x(x h)] \right ) \ee which
is no longer just $v_{\hat{x}}=Kx$ as used in Sec.
\ref{exponential}, but takes into account all the terms entering
the tangential velocity jump there [Eq. (\ref{veljump})].

We now present a general solution when the order $\varepsilon$ in
Eq.(\ref{expandedevolution}) is neglected. Any given initial
condition $h_0(x)$ evolves following a simple scaling
$h(x,t)=Lh_0(Lx)$ [Eq. (\ref{scalansatz})] with $L=e^{-Kt}$ and
$K=2/(1+A)$. This agrees with the scaling found in Sec.
\ref{scalings} for an initially straight interface. Here, we find
that the result is more general and, as a consequence, any
interface approaches zero at least at infinite time. In this way,
rotation produces pinch-off at infinite time independently of the
initial interface. This is to be compared with the influence of a
gravity jet which only produces a shift in the moving frame.

This leading behavior smooths out the gradients both by lowering
height differences [Eq. (\ref{scalingh})] and by stretching
neighboring points apart [Eq. (\ref{scalingr})]. One would expect
straight filaments to be stable, and the solution to become
increasingly accurate in time. This smoothing effect of the
centrifugal force through the zeroth order term $v_{\hat{x}}=Kx$
presumably competes with the higher order terms: The term in
dimensionless surface tension $B$, for instance, is known to lead
to finite-time pinch-off in the absence of any other force and for
an inviscid outer fluid ($A=1$), for certain initial conditions
\cite{Almgren95, Almgren96,Goldstein98}. The analysis of the
interplay between rotation and surface tension and their competing
effects deserves an in-depth separate analysis that will be
carried out elsewhere.

However, we can address in this framework the systematic
observation of pinch-off events in experiments and simulations for
low viscosity contrasts $A=0$--0.5 and the lack of them for high
viscosity contrast. This fact cannot be explained by the two first
terms in the r.h.s of (\ref{expandedevolution}). If such
singularities and differences are contained in the Hele-Shaw
dynamics as the simulations seem to indicate, they should be
linked to the remaining non local term of order $\varepsilon$.
Although this term does indeed play a role for $A<1$, we have seen
that it enters at the same order than the surface tension term.
The question is then how this, in principle, higher order term can
affect the general asymptotic scaling of filaments which typically
become straight for any viscosity contrast.

The answer is in the {\em non-local} nature of the new term.
Hence, a locally straight filament is still influenced by a
curvature elsewhere. Real filaments are finite, and this means
that the upper and lower interfaces necessarily meet in a highly
curved region (in practice, a droplet at the tip), where the
$\varepsilon$ expansion breaks down. Because this will be felt by
an elsewhere straight filament, the $\varepsilon$ expansion is
much more delicate for low viscosity contrast, and significant
deviations from the scaling of Eqs. (\ref{scalansatz}) or
(\ref{scalingh}) can be expected. The influence of a curved region
decays with distance, so that these deviations should decay for
the central part of a filament as it grows long enough for its
ends to have little effect on it.

Although it is difficult to address the precise effect of the
non-local term analytically, we can gain some insight by
considering the simplest possible situation: a perfectly straight
filament of time-dependent arbitrary width. More precisely, we
consider only an upper straight interface given by $y=h(t)$ in
Fig. \ref{sketch} and the region $y>0,\,x>0$, with no-flux
boundary conditions on the axis $y=0$ and $x=0$, exactly as in our
simulations in Sec. \ref{thinning}. Proposing a stream function
$\psi$ (bi)linear in $x$ and $y$ to solve the full Hele-Shaw
equations [Eqs. (\ref{fbp})], we find
\begin{subequations}
\label{condensador}
\be
\psi_{\rm in} = K_{\rm in} xy, \\
\psi_{\rm out} = h(K_{\rm in}-K_{\rm out}) x  +  K_{\rm out} xy, \\
h=h(t=0)e^{-K_{\rm in}t},\\
K_{\rm in}=K+\frac{(1-A)}{1+A}K_{\rm out},
\ee
\end{subequations}
where we recall that $K=2/(1+A)$.

For $A=1$, we recover $K_{\rm in}=K$ because the inner fluid and
interface dynamics are completely specified and decoupled from the
outer flow. For $A<1$, however, the whole solution depends on the
outer flow, and hence on actual external boundary conditions. For
a semiinfinite filament and no {\em external} breaking of the
translational symmetry, $K_{\rm out}=0$, and the scaling solution
is again valid and exact.

For a finite filament, though, there will necessarily be a
recirculation (not to speak of the flow created by a droplet):
Stream lines going out of the filament at its tip will come back
to different points of the interface, in a loop of the size of the
filament length $R$ itself. The outer flow solution proposed will
no longer be valid, but, for a long enough filament, we can assume
that the solution still holds for $x\ll 1$. Given the fact that
recirculation requires a positive hyperbolic profile close to the
origin (see Fig.\ \ref{stream}), $K_{\rm out}$ must be positive
and therefore $K_{\rm in}>K$, slowly approaching $K$ as the
filament grows longer. This is consistent with the filament
thinning exponent for $A=0$ observed in Fig. \ref{thinningsimul}.

One could speculate that this effect becomes stronger as one
approaches the filament tip (although the flow also looks
different), making the local, effective thinning (and its rate)
increase. This would lead to the formation of a more shallow
region (the neck) and thus initiate the pinch-off. The forming
neck then happens to receive a much higher density of incoming
stream lines that its vicinity (see Fig. \ref{stream}, blowup at
the bottom left), pushing it further down. Surface tension enters
the problem as the filament is locally curved, so that it is not
possible to tell whether the pinch-off itself is due to the
non-local term or to surface tension, or yet to a combination of
the two unless a careful analysis is performed. It is nevertheless
clear that the non-local term does provide the initial departure
from the scaling solution needed to overcome its smoothing effect.

\section{Discussion and conclusions}
\label{discussion}

We have studied viscous fingering in a rotating Hele-Shaw cell
from the point of view of the dynamical approach to pinch-off
singularities. We have found that this is linked to a global
scaling of the pattern: Radial filaments of denser inner fluid
narrow and stretch; they push away the droplet at their tip, and
incoming fingers of less dense, outer fluid approach the rotation
axis to compensate for the flux expelled through the filaments.
For $n$-fold symmetric patterns, the system can approach either a
steady shape in infinite time or a finite-time pinch-off
singularity. The former consists of $n$ convex arches ending in
straight, formally semiinfinite radial filaments of zero width
with a droplet at their tip (at infinity) connecting each pair of
arches and running away with infinite velocity.

Thus, we see that rotation enforces pinch-off as expected, at
least in an infinite time. More precisely, at the lowest order in
thickness variations of a lubrication approximation for the
filaments, they narrow exponentially in time.
Incoming fingers slow down in the same way, whereas the filaments
stretch and the droplets at their tips are centrifuged away also
exponentially.

In the high viscosity contrast limit, $A\to 1$, all these
exponential behaviors have the same time constant, $1/t^*$, where
$t^*$ is the characteristic time scale of the problem, and we have
been able to observe all of them and confirm this value by
phase-field simulations and experiments. For low viscosity
contrasts, the filament thinning has a time constant of
$K=2/[(1+A)t^*]$ in absolute value, whereas the droplet dynamics
remain close to $K'=1/t^*$ as for high viscosity
contrasts. The droplet scaling is thus universal in a first approximation,
i.e., only weakly dependent on any dimensionless control parameter.
Furthermore, we have found it not to be much
affected by whether the droplet is isolated or connected to a
filament. We have derived an exact time-dependent solution
(previously found in the $A\to 1$ limit \cite{tesifrancesc})
for the case of an off-center isolated circular droplet which shows
that this universality is rigorous for isolated circular droplets.

In a second approximation, droplets at
the tip of the filaments are actually
found to scale with a time constant slightly below the theoretical
value $1/t^*$, so that this is always below that of the filament
thinning $2/[(1+A)t^*]$.
Some fraction of the fluid expelled by the filament as
it narrows is hence fed into the droplet. Because the time
constant for the thinning increases above $1/t^*$ as $A$ is
decreased, the lower $A$ the more flux the droplets receive, as
observed in both experiments and simulations.

For experiments with two liquids (low $A$), comparison with the exact
solution for an isolated circular droplet and extensive simulations
have shown the time scale of the dynamics $t^*$ to be lower
than its characteristic value
$t^*=12(\mu_{\rm in}+\mu_{\rm out})/
(b^2\Omega^2\Delta\rho)$. We explain this by an effective
lower viscosity of the inner fluid $\mu_{\rm in}$, which comes
from Darcy's law corrected to take into account the presence
of a wetting layer of the outer liquid. This results in an increased
inner liquid mobility.

Although rotation causes infinite-time pinch-off at the lowest
order in the variations of the filament thickness, both the thickness
decay and the stretching apart of neighboring points smooth out
thickness variations. Rotation also enters the other next order term:
a non-local term proportional to $1-A$. Because this term is
non-local and the interface is necessarily highly curved in the
droplets at the end of the filaments, it can overcome the
stabilizing effect of the lowest-order contribution of rotation,
and thus initiate finite-time pinch-off. The role of surface
tension and the interplay with the local and non-local terms
associated with rotation remain interesting open questions.

In any case, this non-local term can account for the differences observed in
both experiments and simulations, where finite-time pinch-off
singularities appear systematically for low values of $A$ but not
for high values of $A$. Three-dimensional effects like wetting do
not seem to account for the initiation of pinch-off, since
simulations are free of them and nevertheless reproduce well the
experimental morphologies and the pinch-off phenomenon itself.
Corrections due to the non-local term are also
the main reason for obtaining a less clear
exponential behavior for low viscosity contrasts.

Note that, for the case of a channel under gravity or pressure-driven
injection, this non-local term still arises, although it has another form.
Therefore, it may also explain the enhancement
of finite-time pinch-off observed for low $A$ in these well known cases.

\section{Acknowledgments}
R. F. acknowledges financial support through a European Community
Marie Curie fellowship. R. F. and J. C. also thank the Benasque
Science Center for hospitality and the European Union high-level
scientific conference and the NATO Advanced Studies Institute
programs for support during  a summer school at which part of this
work was carried out. Financial support from Ministerio de Ciencia
y Tecnologia (Spain) under project BQU2003-05042-C02-02 and from
European Commission's Research Training Network HPRN-CT-2002-00312
(PHYNECS) are acknowledged.


\begin{references}

\bibitem{scalefree}
J. Keller and M. J. Miksis, SIAM J. Appl. Math. {\bf 43}, 268 (1983).

\bibitem{Eggers97}
J. Eggers, Rev. Mod. Phys. {\bf69}, 865 (1997).

\bibitem{Science00}
M. Moseler and U. Landman, Science {\bf289}, 1165 (2000).

\bibitem{Eggers02}
J. Eggers, Phys. Rev. Lett. {\bf89}, 084502 (2002).

\bibitem{Oron97}
A. Oron, S. H. Davis, and S. G. Bankoff, Rev. Mod. Phys. {\bf69},
931 (1997).

\bibitem{Zhang99}
W. W. Zhang and J. R. Lister, Phys. Fluids {\bf11}, 2454 (1999).

\bibitem{Lee02} For a
way to model scales below the cell gap, see
H-G. Lee, J.S. Lowengrub, and J. Goodman, Phys. Fluids {\bf14},2
492 (2002), Phys. Fluids {\bf14}, 514 (2002) .

\bibitem{Goldstein93}
R.E. Goldstein, A.I. Pesci, and M.J. Shelley, Phys. Rev. Lett.
{\bf70}, 3043 (1993).

\bibitem{Duppont93}
T.F. Duppont, R.E. Goldstein, L.P. Kadanoff, and S-M. Zhou, Phys.
Rev. E {\bf47}, 4182 (1993).

\bibitem{Almgren95}
R. Almgren, Phys. Fluids {\bf 8},2 344 (1995).

\bibitem{Almgren96}
R. Almgren, A. Bertozzi, and M. Brenner Phys. Fluids {\bf 8},6
1356 (1996).

\bibitem{Constantin93}
P. Constantin, T.F. Duppont, R.E. Goldstein, L.P. Kadanoff, M.J.
Shelley, and S-M. Zhou, Phys. Rev. E {\bf47}, 4169 (1993).

\bibitem{Goldstein95}
R.E. Goldstein, A.I. Pesci, and M.J. Shelley, Phys. Rev. Lett.
{\bf75}, 3665 (1995).

\bibitem{Goldstein98}
R.E. Goldstein, A.I. Pesci, and M.J. Shelley, Phys. Fluids
{\bf10}, 2701 (1998).

\bibitem{Sierou03}
A. Sierou and J. R. Lister, J. Fluid Mech. {\bf 497}, 381 (2003).

\bibitem{st}
P.G. Saffman and G.I. Taylor,
Proc. R. Soc. London A, {\bf 245}, 312 (1958).

\bibitem{maher}
J. V. Maher, Phys. Rev. Lett., {\bf 54}, 1498 (1985).

\bibitem{footnote0}
We recall here that injection and gravity are completely equivalent,
since both can be reparametrized into a unique driving force.

\bibitem{Alvalow03}
E. Alvarez-Lacalle, J. Casademunt, and J. Ort\'{\i}n, Phys.
Fluids. {\bf 16} 908 (2004).

\bibitem{Alvahigh03}
E. Alvarez-Lacalle, J. Casademunt, and J. Ort\'{\i}n, (preprint).

\bibitem{prewet}
As demonstrated in \cite{Alvahigh03}, a thin uniform layer of the oil
wetting the glass plates is necessary to reproduce the numerical results
obtained with standard boundary conditions. This is done in practice
by generating an advancing stable annulus of the same oil
before the experiment is performed (see \cite{Carrillo-99}).

\bibitem{Paune-thesis}
E. Paun\'e, \it Interface dynamics in two--dimensional viscous
flows \rm Ph. D. thesis, Universitat de Barcelona (2002).

\bibitem{geometrical}
Z. Csah\'ok, C. Misbah and A. Valance,
Physica D {\bf 128}, 87 (1999).

\bibitem{tesifrancesc}
F.X. Magdaleno, Ph. D. thesis, Universitat de Barcelona (2000).

\bibitem{elastica}
E. Alvarez-Lacalle, J. Casademunt, and J. Ort\'{\i}n, Phys. Rev.
Lett. {\bf92}, 054501 (2004).

\bibitem{Tryggvason85}
G. Tryggvason and H. Aref, J. Fluid Mech. \bf154\rm, 287 (1985).

\bibitem{folch00} For a step-by-step derivation (including a kind of
viscosity anisotropy), see
R. Folch, J. Casademunt and A. Hern\'andez-Machado, Phys. Rev. E
{\bf 61}, 6632 (2000).

\bibitem{pf1}
R. Folch, J. Casademunt, A. Hern\'andez-Machado and L. Ram\'{\i}rez-Piscina,
Phys. Rev. E {\bf 60}, 1724 (1999).

\bibitem{pf2}
R. Folch, J. Casademunt, A. Hern\'andez-Machado and L. Ram\'{\i}rez-Piscina,
Phys. Rev. E {\bf 60}, 1734 (1999).

\bibitem{pinch3d}
T. Biben, C. Misbah, A. Leyrat and C. Verdier,
Europhys. Lett. {\bf 63}, 623 (2003).

\bibitem{Park} C.W. Park and G.M. Homsy, J. Fluid Mech.
{\bf 139}, 291 (1984).

\bibitem{Alvarez01}
E. Alvarez-Lacalle, J. Casademunt, and J. Ort\'{\i}n, Phys. Rev. E
{\bf 64}, 016302 (2001).

\bibitem{Carrillo-99}
L. Carrillo, J. Soriano, and J. Ort\'\i n. Phys. Fluids. {\bf 11}
778 (1999).

\end{references}
\end{document}